\documentclass[%
 prx,reprint,groupedaddress,
 amsmath,amssymb,
 aps,
]{revtex4-2}

\usepackage{verbatim}
\usepackage{rotating, booktabs}  
\usepackage{subfigure}
\usepackage{braket}
\usepackage{graphics}
\usepackage{graphicx}
\usepackage{braket}
\usepackage{hyperref}
\usepackage{listings}
\usepackage{color, xcolor}
 \usepackage{adjustbox}
\usepackage{upgreek}
\usepackage{dcolumn}
\newcolumntype{z}[1]{D{.}{.}{#1}}
\usepackage{float} 

 \usepackage{tikz}
 \usepackage{circuitikz}
 \usetikzlibrary{quantikz}
 \usetikzlibrary{decorations.pathreplacing}

 \usetikzlibrary{backgrounds,fit,decorations.pathreplacing}  
 \usetikzlibrary{automata} 
 \usetikzlibrary{circuits.logic.CDH}
 \usetikzlibrary{circuits.logic.US}
 \usetikzlibrary{mindmap}
 \usetikzlibrary{decorations}
 \usetikzlibrary{decorations.pathmorphing}

 \usetikzlibrary{arrows,decorations.pathreplacing}

 \tikzset{meter/.append style={draw, inner sep=10, rectangle, font=\vphantom{A}, minimum width=30, line width=.4,
 		path picture={\draw[black] ([shift={(.1,.3)}]path picture bounding box.south west) to[bend left=50] ([shift={(-.1,.3)}]path picture bounding box.south east);\draw[black,-latex] ([shift={(0,.1)}]path picture bounding box.south) -- ([shift={(.3,-.1)}]path picture bounding box.north);}}}

\usepackage{extpfeil}

\usepackage{wasysym} 

\usepackage{etoolbox}
\patchcmd{\thebibliography}{\section*{\refname}}{}{}{}  
\usepackage{url}
\usepackage{multirow} 
\usepackage[mathscr]{eucal}

\usepackage{shadow,epsf,amsthm,amssymb,amsmath}

\usepackage{algorithm}
\usepackage{algpseudocode}
  
\usepackage[table]{xcolor} 
\definecolor{zcol}{RGB}{200, 255, 200} 
\definecolor{xcol}{RGB}{255, 200, 200} 
\definecolor{c1_dark}{RGB}{0, 60, 130}    
\definecolor{c1_light}{RGB}{80, 160, 240} 

\definecolor{c2_dark}{RGB}{150, 0, 0}     
\definecolor{c2_light}{RGB}{240, 100, 100} 

\definecolor{c3_dark}{RGB}{0, 90, 40}     
\definecolor{c3_light}{RGB}{100, 190, 110} 

\definecolor{c4_dark}{RGB}{170, 80, 0}    
\definecolor{c4_light}{RGB}{250, 160, 60}

\newtheorem{definitionenv}{Definition}
\newtheorem{lemmaenv}[definitionenv]{Lemma}
\newtheorem{theoremenv}[definitionenv]{Theorem}
\newtheorem{corollaryenv}[definitionenv]{Corollary}
\newtheorem{propositionenv}[definitionenv]{Proposition}
\newtheorem{remarkenv}[definitionenv]{Remark}
\newenvironment{remark}{\begin{remarkenv}\rm}{\end{remarkenv}}
\newcommand{\br}{\begin{remark}}
	\newcommand{\er}{\end{remark}}

\newtheorem{exampleenv}{Example}
\newtheorem{app-lemmaenv}[section]{Lemma}

\newenvironment{definition}{\begin{definitionenv}\rm}{\end{definitionenv}}
\newenvironment{lemma}{\begin{lemmaenv}\rm}{\end{lemmaenv}}
\newenvironment{theorem}{\begin{theoremenv}\rm}{\end{theoremenv}}
\newenvironment{corollary}{\begin{corollaryenv}\rm}{\end{corollaryenv}}
\newenvironment{example}{\begin{exampleenv}\rm}{\end{exampleenv}}
\newenvironment{proposition}{\begin{propositionenv}\rm}{\end{propositionenv}}
\newenvironment{app-lemma}{\begin{app-lemmaenv}\rm}{\end{app-lemmaenv}}

\newcommand{\bd}{\begin{definition}}
	\newcommand{\ed}{\end{definition}}
\newcommand{\bl}{\begin{lemma}}
	\newcommand{\el}{\end{lemma}}
\newcommand{\elp}{\hspace*{\fill} $\Box$
\end{lemma}}
\newcommand{\bt}{\begin{theorem}}
\newcommand{\et}{\end{theorem}}
\newcommand{\etp}{\hspace*{\fill} $\Box$
\end{theorem}}
\newcommand{\bc}{\begin{corollary}}
\newcommand{\ec}{\end{corollary}}
\newcommand{\ecp}{\hspace*{\fill} $\Box$
\end{corollary}}

\newcommand{\be}{\begin{example}}
\newcommand{\ee}{\end{example}}
\newcommand{\eep}{\hspace*{\fill} $\Box$
\end{example}}
\newcommand{\bp}{\begin{proposition}}
\newcommand{\ep}{\end{proposition}}
\newcommand{\epp}{
\end{proposition}}

\usepackage{xspace}

\usepackage[colorinlistoftodos,prependcaption]{todonotes}
\usepackage{xargs}

\newcommandx{\yellownote}[2][1=]{\todo[inline,linecolor=yellow,backgroundcolor=yellow!25,bordercolor=yellow,#1]{#2}}

\newcommand{\bfa}{ \mathbf{a}}
\newcommand{\bfb}{ \mathbf{b}}
\newcommand{\bfc}{ \mathbf{c}}
\newcommand{\bfx}{ \mathbf{x}}
\newcommand{\bfy}{ \mathbf{y}}
\newcommand{\bfe}{ \mathbf{e}}

\newcommand{\bfg}{ \mathbf{g}}
\newcommand{\bfh}{ \mathbf{h}}

\newcommand{\cE}{{\cal E}}

\newcommand{\cG}{{\cal G}}

\newcommand{\cN}{{\cal N}}
\newcommand{\cS}{{\cal S}}

\newcommand{\cD}{{\cal D}}

\newcommand{\cP}{{\mathcal P}}

\newcommand{\CDR}{{\text{DR}}}

\newcommand{\wt}[1]{{\mathrm{wt}\left\{ #1\right\}  }}
 
\allowdisplaybreaks 

\newcommand{\twirl}{\ensuremath{\mathrm{Twirl}}}
\newcommand{\tr}{\ensuremath{\mathrm{tr}}}

\begin{document}

 \title{Fault-Tolerant Quantum Error Correction for Constant-Excitation Stabilizer Codes under Coherent Noise}%

\author{Ching-Yi Lai}
\email{cylai@nycu.edu.tw}
\affiliation{Institute of Communications Engineering, National Yang Ming Chiao Tung University, Hsinchu 30010, Taiwan}
\affiliation{National Center for Excellence in Quantum Information Science and Engineering}

\author{Pei-Hao Liou}
\affiliation{Institute of Communications Engineering, National Yang Ming Chiao Tung University, Hsinchu 30010, Taiwan}
\author{Yingkai Ouyang}
\affiliation{School of Mathematical and Physical Sciences, University of Sheffield, Sheffield, UK}
\date{\today}
 
\begin{abstract}
Collective coherent (CC) noise poses challenges for fault-tolerant error correction (FTEC), as it is not captured by conventional stochastic noise models. Constant-excitation (CE) codes are inherently immune to CC errors, but a fault-tolerant framework for operating these codes under circuit-level noise has not yet been established.
{Here, we develop an FTEC framework for CE CSS codes based on dual-rail concatenation.} We show that conventional transversal CNOT gates violate the CE constraint and develop CE-preserving logical CNOT gates together with modified Shor- and Steane-type syndrome extraction schemes using zero-controlled NOT gates and CE-compatible ancilla states. We further develop an extended stabilizer simulation algorithm that tracks both stochastic and CC noise. Using this framework, we identify small distance-3 CE CSS codes {demonstrate that the $[[14,1,3]]$ code maintains robust performance under coherent noise. 
	Our results establish a fault-tolerant framework for CE codes under circuit-level noise and demonstrate their potential for quantum processors affected by CC noise.}

\end{abstract}





\maketitle
  
\section{Introduction}

In practical quantum computing systems, noise limits the reliability of quantum computation. Fault-tolerant quantum computation (FTQC)~\cite{DS96,Got97,AB97} provides a framework for performing reliable quantum computation despite faulty gates and environmental decoherence. A central concept in FTQC is the fault-tolerant threshold~\cite{knill1998resilient,AGP06,Got14c}, namely the maximum tolerable error rate below which arbitrarily long quantum computations become feasible using quantum error-correcting codes. Consequently, considerable effort has been devoted to reducing the overhead of FTQC and improving logical error rates through optimized fault-tolerant protocols, quantum error-correcting codes, and decoding algorithms~\cite{Bra+24,KL24}.
Threshold estimation inevitably involves a trade-off between model simplicity and physical accuracy. Stochastic independent noise models are widely adopted because they enable tractable analyses. However, such models often fail to capture realistic hardware noise, and practical methods for analyzing more accurate noise models remain limited.

Collective coherent (CC) errors arise when groups of qubits undergo correlated unitary rotations due to shared control infrastructure or global noise sources. Such errors occur in various quantum computing platforms. For example, in trapped-ion systems, CC errors can result from laser amplitude or phase miscalibrations that induce uniform over-rotations across multiple ions~\cite{SPJ+11,GTL+16}. In superconducting qubit systems, similar collective over-rotations can be caused by shared microwave control lines, global oscillator phase drift, or residual ZZ interactions~\cite{MWS+16,KTC+19}. Furthermore, quantum systems whose timing is governed by a master clock, such as trapped-ion~\cite{monroe2013scaling}, superconducting~\cite{oliver2013materials}, and semiconductor qubits~\cite{reilly2015engineering}, are susceptible to CC errors arising from clock instability~\cite{ball2016role}.

Simulating fault-tolerant error correction (FTEC) under coherent noise presents a fundamental challenge, as CC errors are non-Clifford. This precludes the use of efficient stabilizer-based simulation techniques~\cite{GK98,PhysRevA.70.052328-aaronsongottesman} and requires tracking non-Clifford quantum states, which is generally computationally intractable~\cite{PRXQuantum.2.010345-magic}. The difficulty remains even when CC errors are described by a simple collective unitary of the form
$\exp(-i\theta \sum_j Z_j)$, where $\theta$ is a rotation angle and $Z_j$ denotes a phase operator on the $j$th qubit.
Although techniques such as randomized compiling~\cite{PhysRevX.11.041039} and dynamical decoupling~\cite{PhysRevLett.90.037901-robust-DD,souza2012robust,PhysRevLett.95.180501-FTDD} can transform coherent errors into effective stochastic noise, these approaches generally do not exploit the underlying structure of coherent errors and may introduce additional stochastic error contributions.
{
Coherent errors in surface codes have been investigated using efficient simulation techniques under perfect syndrome extraction and phenomenological noise models with measurement errors~\cite{BEK+18,MA23}. More recently, syndrome-derived detector error models have been explored for estimating and decoding coherent noise in QEC experiments~\cite{TB25}.
}

{
	Constant-excitation (CE) stabilizer codes~\cite{Ouy21,HLRC22,CL25}, which are immune to CC errors, provide an approach for mitigating CC error accumulation.
	However, exploiting this property under circuit-level noise requires FTEC protocols compatible with the CE constraint, since CC errors are repeatedly introduced throughout syndrome extraction circuits.
	Conventional FTEC protocols rely on transversal CNOT gates~\cite{Shor96}, which are widely used for logical operations and syndrome extraction. However, transversal CNOT gates can map CE states outside the CE subspace, violating the excitation-preserving property of CE codes.
}

We address these challenges through three main contributions:

First, we establish the framework of CE CSS stabilizer codes and characterize their properties under CC errors. In particular, we show that the $[[12,1,3]]$ code encoding one logical qubit and the $[[14,3,3]]$ code encoding three logical qubits are the two smallest distance-3 CE CSS codes constructed via dual-rail concatenation. These codes preserve the excitation constraint while correcting arbitrary single-qubit errors.
 
Second, we develop FTEC protocols compatible with CE CSS codes under circuit-level noise. 
{By designing excitation-preserving syndrome extraction circuits and a dedicated control flow, our protocols enable FTEC in the presence of both stochastic and CC errors for distance-3 CE CSS codes.}

{Third, we develop simulation algorithms for the proposed CE FTEC protocols and evaluate their performance under circuit-level noise with interleaved CC errors. We show that conventional CSS codes, including the rotated surface code and the Steane code, suffer from increasing logical error rates as the coherent rotation angle grows, due to the accumulation of CC errors over syndrome extraction cycles. In contrast, the $[[14,1,3]]$ CE CSS code maintains stable logical error rates under the same noise model, demonstrating its robustness against CC errors.
}


 \section{The Potential of CE Codes for Coherent Noise}
 \label{sec:potential}

	The CC error
	\[
	e^{-i \theta \sum_j Z_j}
	=
	\prod_j e^{-i \theta Z_j},
	\]
	can be viewed as a product of single-qubit coherent rotations, where
	$e^{-i\theta P}$, for a Pauli operator
	$P\in\{I,X,Y,Z\}^{\otimes n}$,
	is referred to as a \emph{coherent rotation}.

A \emph{constant-excitation} (CE) state of length \( n \) and weight \( w \) is  a linear combination of $n$-qubit computational basis states with fixed Hamming weight \( w \), and can be written as
$\ket{\psi} = \sum_{{\bf x} \in \{0,1\}^n,\ \wt{{\bf x}} = w} \alpha_{\bf x} |{\bf x}\rangle$. Here, \( \wt{{\bf x}} = x_1+\dots + x_n \) denotes the Hamming weight of \( {\bf x} = (x_1,\dots,x_n) \), and \( \alpha_{\bf x} \in \mathbb{C} \). We refer to such a state as a $w$-CE 
state. A $w$-CE code is any quantum code whose codespace comprises $w$-CE states. As eigenstates of the CC error $e^{-i \theta \sum_j Z_j}$, CE codes are inherently immune to such errors \cite{Ouy21}.  
We provide further details on CE CSS stabilizer codes in Appendix~\ref{app:CE_codes}.

Current designs of fault-tolerant quantum computations do not use CE codes. This necessitates mitigating CC noise with methods such as randomized compiling \cite{PhysRevX.11.041039} or dynamical decoupling \cite{PhysRevLett.90.037901-robust-DD,souza2012robust,PhysRevLett.95.180501-FTDD}, and subsequent twirling of the resultant noise~\cite{ESM+07}. 

{
To motivate our discussion, we consider the following noise model: $\mathcal M(\rho)  =    \cD_p^{\otimes n}(U\rho U^\dagger)$, 
which consists of the composition of a unitary CC noise channel
 $U = \exp( i \theta \sum_{j}Z_j ) $, followed by
an $n$-qubit depolarizing channel $\mathcal D_p^{\otimes n}$.
The  parameter
$\theta $ quantifies the strength of coherent noise, 
 The single-qubit depolarizing channel is given by
$\mathcal D_p(\rho) = (1-p) \rho + (p/3) (X \rho X^\dagger + Y \rho Y^\dagger + Z \rho Z^\dagger)$, where $X$, $Y$ and $Z$ are Pauli matrices, and  $p$ is the depolarizing error rate.

To illustrate the impact of coherent errors, we consider a simplified model where the coherent component is Pauli-twirled~\cite{ESM+07}:  
\begin{align}
	\tilde{\mathcal M}(\rho)  =    \cD_p^{\otimes n}( \cP_{\textrm{coh}}^{\otimes n}(\rho)),\label{eq:FT-noise-model2}
\end{align}
where $\cP_{\textrm{coh}}(\rho)= \cos^2\theta \rho + \sin^2\theta Z \rho Z$ as provided in Appendix~\ref{app:twirling}.
Therefore,  $\tilde{\mathcal M}= (\cD_p\circ\cP_{\textrm{coh}})^{\otimes n}$ and 
$\cD_p\circ\cP_{\textrm{coh}} = \cP_q$, 
where $\mathcal P_q$ denotes the resulting single-qubit Pauli channel
 with rates $q_I=1-p-\sin^2{\theta}+\frac{4}{3}p\sin^2{\theta} $, $q_X=q_Y=\frac{p}{3}$
and $q_Z=\sin^2{\theta}+\frac{p}{3}-\frac{4}{3}p\sin^2{\theta} $.
This shows that coherent errors, after Pauli twirling, contribute an additional stochastic \(Z\)-error component, thereby increasing the effective \(Z\)-error rate when \(p<3/4\).
}

{
Previous work has shown that coherent rotation errors can significantly degrade the performance of conventional surface code error correction, particularly when the rotation angle becomes sufficiently large~\cite{BEK+18}. 
Under Pauli twirling, such coherent rotations are converted into an effective stochastic Pauli channel with a $Z$ error rate
\[
q_Z=\sin^2{\theta}+\frac{p}{3}-\frac{4}{3}p\sin^2{\theta},
\]
which can exceed the error threshold of conventional quantum codes even when the depolarizing error rate $p$ alone remains below threshold.	
This vulnerability of conventional quantum error-correcting codes motivates the development of CE CSS codes, which are designed to tolerate coherent rotation errors without converting them into detrimental logical errors.
}


We note that all theorem and lemma proofs are provided in the appendix.


\section{Quantum circuit  model}

We assume that each circuit location—whether a gate, measurement, state preparation, or idle qubit—takes the same amount of time to execute. Consequently, we decompose a quantum circuit into discrete gate depths, where each depth corresponds to a uniform time interval.

For error correction, in addition to single-qubit measurements in the $Z$ or $X$ basis, we assume the availability of the following Clifford gates.
\begin{align*}
CX_{1,2}=&	\ket{0}\bra{0}\otimes I +\ket{1}\bra{1}\otimes X,\\
CZ_{1,2}=& \ket{0}\bra{0}\otimes I+ \ket{1}\bra{1}\otimes Z,\\
C_0 X_{1,2}\triangleq&  	\ket{0}\bra{0}\otimes X +\ket{1}\bra{1}\otimes I,\\
C_0Z_{1,2}\triangleq& \ket{0}\bra{0}\otimes Z+ \ket{1}\bra{1}\otimes I.
\end{align*}
Here, $CX$ and $CZ$ denote the standard controlled-NOT and controlled-$Z$ gates, respectively, while $C_0X$ and $C_0Z$ represent the zero-controlled counterparts.
  We require the implementation of these gates to be as single native gates rather than as gate sequences to prevent CC errors from introducing unintended coherent rotation errors. 

\begin{figure}[htbp]{
\includegraphics[width=0.99\columnwidth]{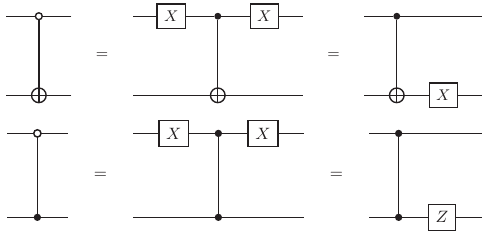}
	}
	\caption{Zero-controlled NOT gate and zero-controlled $Z$ gate. }
	\label{fig:CDR}
\end{figure}

We can also realize the zero-controlled NOT gate $C_0 X_{1,2}$ via the decomposition $X_2 \cdot CX_{1,2}$ as shown in Fig.~\ref{fig:CDR}. Recent advances in superconducting qubit control have demonstrated the feasibility of implementing high-fidelity two-qubit gates--including composite gates like the zero-controlled NOT--using optimized microwave pulse sequences and shaped control fields \cite{Sheldon2016,EW14,HG14}. 
These techniques leverage optimal control theory and cross-resonance effects to achieve robust operations with reduced error rates and shorter gate times, demonstrating the experimental feasibility of implementing such composite two-qubit operations.

In this paper, we consider a realistic circuit-level noise model interleaved with CC noise. Every component of the quantum circuit is subject to faults.  Assuming a physical error rate $p$, the noise model includes the following error mechanisms:
\begin{itemize}
\item Each single-qubit gate or ancilla state preparation is followed by a Pauli error ($X$, $Y$, or $Z$), each occurring with probability $p/3$.
\item Each idle location undergoes a Pauli error ($X$, $Y$, or $Z$), each occurring with probability $p/3$.

	\item Each two-qubit gate is followed by a non-identity two-qubit Pauli error, uniformly selected from the 15 nontrivial two-qubit Pauli operators, each occurring with probability $p/15$.
	\item The outcome of a single-qubit measurement is flipped with probability $2p/3$.
 \end{itemize}

 In addition to circuit location faults, we consider coherent errors in the $Z$ direction, modeled by a unitary operator of the form $e^{-i\theta Z}$, where $\theta$ is a real-valued rotation angle. 
 For an $n$-qubit system, we assume that all qubits experience the same coherent rotation, resulting in a CC error described by
 \[
 \exp\left(-i\theta\sum_j Z_j\right).
 \]
 In our noise model, a layer of CC error is applied after each circuit-depth layer. 
 Here, a circuit-depth layer consists of parallel two-qubit gates or measurements that can be performed simultaneously.

 {
 	In our layered circuit model, a CC error layer is inserted after each circuit depth and 	
 	acts collectively on all qubits in the corresponding circuit block involved in the corresponding circuit layer, including data qubits and ancilla qubits used for syndrome extraction. Idle qubits are also included in the CC error layer.
 }
  For illustration, Fig.~\ref{fig:412_circuit} shows a modified syndrome extraction circuit for the $[[4,1,2]]$ CE code, where CC error layers are inserted after each two-qubit gate layer of the syndrome extraction circuit.
 This layered structure enables the analysis of the accumulation and propagation of coherent errors throughout the fault-tolerant circuit.

\begin{figure}[htbp] {
	\centering
}
 
\centering
\tikzset{
	noisy/.style={fill=yellow,draw=red,line width=2pt}
}
\resizebox{1.\columnwidth}{!}{
	\definecolor{color3}{RGB}{44, 160, 44}    
	\begin{quantikz}[row sep=0.3cm, column sep=0.2cm, ampersand replacement=\&]
				 \&    \&   \& \& \& \& \&  \& \& \& \& \& \& \& \& \&     \\
		\lstick[4]{data}\& \ctrl{4}  \& \qw \& \qw\& \qw  \& \gate[8,style={noisy},label style=cyan]{\text{CC}}\&\qw \& \qw \& \qw \& \qw \& \targ \& \qw \& \qw \& \qw \& \qw \& \gate[8,style={noisy},label style=cyan]{\text{CC}} \& \qw \& \qw \& \qw \&     \\
		\& \qw \& \ctrl{4} \&\qw \& \qw \& \qw\&\qw\&\qw \&\qw\&\qw\&\qw\&  \targ\&\qw\&\qw\&\qw \& \qw\&\qw \& \qw\&\qw\& \\
		\& \qw \& \qw \& \ctrl{4} \& \qw\& \qw\&\qw\&\qw  \&\qw\& \qw\&\qw \& \qw\&   \targ\&\qw\&\qw\&\qw\&\qw \& \qw\&\qw\& \\
		\& \qw \& \qw \& \qw\& \ctrl{4} \& \qw\& \qw\&\qw\& \qw\& \qw\&\qw \& \qw\&  \qw\&   \targ\&\qw\&\qw\&\qw \& \qw\&\qw\&\\
		\lstick[2]{ancilla}\& \ctrl{-1} \& \qw   \& \qw\& \qw\&  \& \meterD[style={draw=blue,{fill=blue!10}}]{M_X}\&\&\&\lstick[4]{ancilla}\&\ctrl{-4}\&\qw\&\qw\&\qw\&\qw\&\meterD[style={draw=color3,{fill=color3!10}}]{M_X}\\
		\& \qw \& \octrl{-1}\& \qw\& \qw \&   \qw \& \meterD[style={draw=blue,{fill=blue!10}}]{M_X}\&\&\&\&\qw\&\octrl{-4}\&\qw\&\qw\&\qw\&\meterD[style={draw=color3,{fill=color3!10}}]{M_X} \& \& \&  \\
		\lstick[2]{ancilla}\& \qw \& \qw  \& \ctrl{-1}\& \qw\& \qw \& \meterD[style={draw=red, {fill=red!10}}]{M_X}\&\&\&\&\qw\&\qw\&\ctrl{-4}\&\qw\&\qw\&\meterD[style={draw=color3,{fill=color3!10}}]{M_X} \& \& \&\\
		\& \qw \& \qw \& \qw\& \octrl{-1} \& \qw\& \meterD[style={draw=red,{fill=red!10}}]{M_X} \& \&\& \& \qw\&\qw\&\qw\&\octrl{-4}\&\qw\&\meterD[style={draw=color3,{fill=color3!10}}]{M_X} \& \& \&
	\end{quantikz}
}

	\caption{Modified Shor syndrome extraction circuit for the $[[4,1,2]]$ CE code.
	The two $Z$-type stabilizer measurements are shown in red and blue, while the $X$-type stabilizer measurement is shown in green.	
Zero-controlled gates are indicated by open circles on the control qubits.
 	Vertical boxes represent layers of CC errors.
The meter labeled $M_X$ denotes a measurement in the $X$ basis.
	The ancilla states are omitted.
{The ancillas are not reused, and at the second depth, the measurements of the ancillas and the $CX/C_0X$ gates on other qubits are performed simultaneously.
At depth one, four two-qubit gates are implemented on the data block and two ancilla blocks. 
After this gate layer, a CC error layer
$\exp\left(-i\theta\sum_{j=1}^{8} Z_j\right)$
is applied collectively to all data qubits together with the ancilla qubits associated with this circuit depth.}
	}
	\label{fig:412_circuit}
\end{figure}

\section{Structure of CE stabilizer codes}




A $w$-CE stabilizer code is a stabilizer code whose codespace consists entirely of $w$-CE states. Various structural properties of CE stabilizer codes have been analyzed in~\cite{HLRC22}. We highlight the following  properties:

\bl
 
 Consider an $n$-qubit $w$-CE stabilizer code defined by a stabilizer group $\cS$. Then, (1) $(-1)^w Z^{\otimes n} \in \mathcal{S}$ and (2) {for any stabilizer or logical operator,
 the total number of its $X$ and $Y$ components is even.}
 
\el
The first property holds because any $w$-CE state is an eigenvector of $Z^{\otimes n}$. The second follows from the requirement that all stabilizers and logical operators must commute with $Z^{\otimes n}$.

 Any $[[n,k,d]]$ stabilizer code transforms into a $[[2n,k,d' \geq d]]$ CE code after concatenation with the dual-rail   code~\cite{Ouy21} \footnote{A slightly more general construction is presented in~\cite{HLRC22}.}.
Here, the dual-rail code~\cite{KLM01} is  defined by the Clifford encoder:
\begin{align}
(a\ket{0}+b\ket{1})\ket{0} \xmapsto{C_0X}
 a\ket{01}+b\ket{10}, \label{eq:UDR}
\end{align}
which maps a single qubit and an ancilla in $\ket{0}$ into a two-qubit state with one excitation.
This encoder consists of a single zero-controlled gate $C_0X$, as shown in Fig.~\ref{fig:CDR}, with the data qubit serving as the control.
In the stabilizer formalism, the dual-rail codespace is 
stabilized by $ -Z\otimes Z,$ and has the logical operators
{
 \begin{align}
 {X}^\CDR=&X\otimes X, \quad
 {Y}^\CDR=Y\otimes X, \quad
 {Z}^\CDR=Z\otimes I
 \label{eq:ZI}.
 \end{align} 
}
 
We describe concatenation with a dual-rail code using the map $\tau$ that takes an $n$-qubit Pauli to a $2n$-qubit Pauli operator, where
{
	\begin{align}
\tau\left(\bigotimes_{i=1}^n M_{i}\right) =\bigotimes_{i=1}^n {M}_{i}^\CDR, 
\end{align}
}
and $M_{i}\in\{I,X,Y,Z\}$ is any single-qubit Pauli.
For example, $\tau(I\otimes X\otimes Y\otimes Z)=I I\otimes X X\otimes Y X\otimes Z I$. 


We formulate the construction of dual-rail-concatenated   CE codes using the stabilizer formalism, as proposed in~\cite{Ouy21}, below.

\bt \label{thm:construction}
Let $\cS=\langle g_1,\dots,g_{n-k}\rangle \subset \cG_n$ be a stabilizer group that defines an $[[n,k,d]]$ stabilizer code.
Define
\begin{align}
\cS'=\Big\langle &\tau(g_i),i=1,\dots,n-k; \notag\\
 &-Z_{2j-1}Z_{2j}, j=1,\dots,n  \Big\rangle.
\end{align}
Then $\cS'$ defines a $[[2n,k,d'\geq d]]$ $n$-CE stabilizer  code.  
Moreover, if $\cS$ defines a CSS code, then so does $\cS'$.
In addition, the symplectic partners of the stabilizers $-Z_{2j-1}Z_{2j}$ for $j = 1, \dots, n$  are given by $X_{2j}$. Each $X_{2j}$ anticommutes with $-Z_{2j-1}Z_{2j}$ and commutes with all other stabilizer generators of $\cS'$.
\et

{
	Since the dual-rail construction introduces $n$ additional $Z$-type stabilizers, it is advantageous to start with an asymmetric CSS code with fewer $Z$-type stabilizers.
 Such a code has a weaker $X$-error correction capability but a stronger relative $Z$-error correction capability. This is because concatenation with the dual-rail code enhances the ability to correct/detect $X$ errors while preserving the ability to correct/detect $Z$ errors, which provides a way to improve the distance and code rate of the resulting dual-rail-concatenated CE code. 
}

In the following, we investigate small distance-3 CE stabilizer codes.
The smallest quantum code that can correct a single-qubit error is the $[[5,1,3]]$ code~\cite{BDSW96}, which yields a $[[10,1,3]]$ CE stabilizer code via Theorem~\ref{thm:construction}.
Restricting our attention to CSS codes, we first establish the following nonexistence results.
\bl \label{non-existence of 8,9 CE CSS}
When $n\le 9$, CE CSS codes $[[n,1,3]]$ do not exist.
\el

\bl \label{non-existence-10CE}
A $[[10,1,3]]$ CE CSS code cannot be obtained from the dual-rail concatenation in Theorem~\ref{thm:construction}.
\el

 Although the second lemma does not rule out the existence of a $[[10,1,3]]$ CE CSS code, no such construction is currently known.
{The $[[7,1,3]]$ Steane code is the smallest distance-3 CSS code, and its dual-rail concatenation produces a $[[14,1,3]]$ CE CSS code.}
 We further obtain the following CE CSS code constructions.

\bt \label{thm-6-CE-12-14construction}
The following CE CSS codes exist:
\begin{enumerate}
	\item A $[[12,1,3]]$ CE CSS code.
	\item A $[[14,3,3]]$ CE CSS code.
\end{enumerate}
\et

%
%
%

These two codes are the smallest distance-3 CE CSS codes constructed via Theorem~\ref{thm:construction}, encoding one and multiple logical qubits, respectively.

\section{Fault-Tolerant Error Correction for CE CSS Codes}

 \subsection{Weak Fault Tolerance}
 
{
The fault-tolerance properties of quantum error correction protocols have been extensively studied in the literature~\cite{AGP06,Got10,CB18,TPB23}. Throughout this paper, we adopt a weak notion of fault-tolerance error correction~\cite{TPB23} based on transversality~\cite{Got10}.

	\bd (Weak FTEC)\label{def:weak_FTEC} An error correction protocol is said to be \textit{$t$-fault-tolerant} if it satisfies the following two conditions.
\begin{enumerate}
	\item[(a)] 	For any input codeword with a Pauli error of weight $e_1$, and for any set of $e_2$ faults occurring during the protocol such that $e_1+e_2\le t$, the output state is correctable by an ideal decoder.
	\item[(b)]  
	For any input codeword with a Pauli error of weight $e_1$, and for any set of $e_2$ faults occurring during the protocol such that $e_1+e_2\le t$, the output state differs from a valid codeword by a Pauli error of weight at most $e_2$.
\end{enumerate} 

\ed

We say that a two-qubit gate is transversal if each qubit in one system interacts with at most one qubit in the other system. Throughout this paper, all multi-qubit interactions are transversal, while all remaining operations are single-qubit gates or measurements. This prevents a single fault from propagating into multiple data-qubit errors within a code block.

}

 \subsection{Basic Identities}
 
{While CE codes are immune to CC errors, they remain vulnerable to  coherent errors acting on subsets of their qubits.
Hence, the study of fault-tolerance for CE stabilizer codes requires understanding how such {coherent rotations} propagate through quantum gates, as characterized by the following lemma.
}

\begin{lemma} The following identities hold: \label{lemma:CE_identity}
	\begin{align}
		&e^{-i\theta Z} X=  X e^{i 2\theta Z}  e^{-i\theta Z}. \label{eq:CE1}\\
		&CX_{1,2} e^{-i\theta Z_2}  =   e^{-i\theta Z_1Z_2} CX_{1,2}.\label{eq:CE2}\\
		&{C_0 X_{1,2} e^{-i\theta Z_2}  =   e^{i\theta Z_1Z_2} C_0 X_{1,2}.}\label{eq:CE3}\\
		&CX_{1,2} e^{-i\theta Z_1}  =   e^{-i\theta Z_1} CX_{1,2}.\\
		&CZ_{1,2} e^{-i\theta Z_1}  =   e^{-i\theta Z_1} CZ_{1,2}.
	\end{align}
	
		%
		%
\end{lemma}


The identity (\ref{eq:CE1}) implies $e^{-i\theta \sum_\ell Z_\ell} X_j=  X_j e^{i 2\theta Z_j}  e^{-i\theta \sum_{\ell}Z_\ell},$ 
indicating that an $X$ error passing through a layer of  CC errors acquires an additional {single-qubit coherent rotation}.
{The identity~(\ref{eq:CE2}) shows that a single-qubit Z coherent rotation propagating through the target qubit of a CNOT gate becomes a two-qubit ZZ coherent rotation}.

The propagation behavior of coherent errors closely resembles that of standard Pauli errors, which suggests that transversal logical operations remain crucial for the fault-tolerant quantum computation of CE stabilizer codes.

It was shown in \cite{HLRC22} that a CE stabilizer code resides in an invariant subspace under a transversal $e^{i\pi/2^k Z}$ gate.
As a result, such diagonal gates act trivially on the CE code space. Therefore, to implement fault-tolerant logical single-qubit gates, fault-tolerant teleportation gadgets, along with the assistance of magic state distillation, may be required~\cite{GC99, BK05, BZHJL14,LZB17,ZLB18,ZLB+20}.

The transversal CNOT gate plays an important role in fault-tolerant syndrome measurement for conventional stabilizer codes. In particular, for a CSS code, a transversal CNOT gate acts as a logical CNOT if all $X$-type and $Z$-type stabilizer generators have phase $+1$~\cite{Got98}. However, this condition does not hold for CE stabilizer codes constructed via Theorem~\ref{thm:construction}, which inherently include $n$ independent stabilizers with phase $-1$.


Without loss of generality, we assume that $\tau(g_1), \dots, \tau(g_{n-k})$ all have phase $+1$ in Theorem~\ref{thm:construction}.
Since transversal CNOT may introduce undesired phases in CE CSS codes, we introduce the following correction to restore the code space.

\bl \label{prop:CNOT}
A transversal CNOT gate followed by the Pauli correction $\prod_{j=1}^n X_{2j}$ preserves the CE CSS code space and implements the logical CNOT operation.
{Equivalently, the corrected operation can be realized by applying $CX$ and $C_0X$ gates alternately on corresponding qubit pairs.}
\el

\begin{figure}[htbp]
\includegraphics[width=0.4\textwidth]{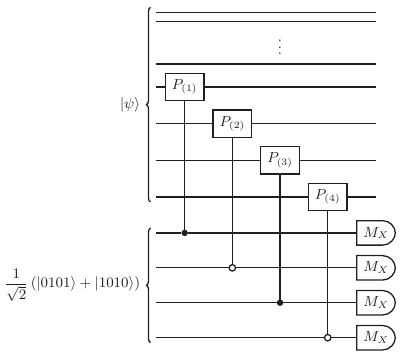}

	\caption{Modified Shor syndrome extraction for a weight-4 stabilizer $P_{(1)}P_{(2)}P_{(3)}P_{(4)}$. 
	} \label{fig:Shor_SE}
\end{figure}


Next, we consider fault-tolerant syndrome extraction for CE codes.
Since ancillary qubits are also subject to coherent errors, we modify the Shor syndrome extraction circuit to satisfy the CE constraint.

In Shor syndrome extraction, 
 a $w$-qubit cat state 
 $
 	\ket{\mathrm{cat}(w)} = \frac{1}{\sqrt{2}}
    (\ket{0}^{\otimes w} + \ket{1}^{\otimes w}
    )
    $
    assists measurement of a stabilizer generator of weight $w$. This cat state
 is stabilized by the operators $X^{\otimes w}, Z_1Z_2, Z_2Z_3, \dots, Z_{w-1}Z_w$. Entanglement in the cat state enables collective measurement of stabilizer generators without disturbing the encoded data.
 
 We adapt Shor's syndrome extraction to CE CSS codes using a $2w$ qubit state
 \begin{align}
 	\ket{\mathrm{cat}^{\mathrm{CE}}(w)} = \frac{1}{\sqrt{2}}\left(\ket{01}^{\otimes w} + \ket{10}^{\otimes w}\right), \label{eq:wCEcat}
 \end{align}
 which we define as a $w$-CE cat state.
 Applying bit-flip operations to half of the qubits in a standard $2w$-qubit cat state efficiently prepares this state:
 \begin{align}
 	\ket{\mathrm{cat}^{\mathrm{CE}}(w)} = (I \otimes X)^{\otimes w} \ket{\mathrm{cat}(2w)}.
 \end{align}
 For low-weight stabilizers with small $w$, the corresponding $w$-CE cat states can be fault-tolerantly prepared using verification circuits, potentially with post-selection~(see, e.g., \cite{EG25}).
 
{
	From Theorem~\ref{thm:construction}, every $X$-type stabilizer in a dual-rail-concatenated CE CSS code has even weight, whereas a $Z$-type stabilizer may have either even or odd weight.
	Denote a weight-$v$ CSS stabilizer by
	$g=(-1)^cP_{(1)}P_{(2)}\cdots P_{(v)}$,
	where $c\in\{0,1\}$ and $P_{(j)}$ denotes the Pauli operator on the $j$-th nontrivial qubit in the support of $g$.
	Since $g$ is a CSS stabilizer, $P_{(j)}=P$ for all $j=1,\ldots,v$,
	where $P\in\{X,Z\}$.
}
 
 Our modified extraction method is presented in Algorithm~\ref{alg:modified_shor}.
  Note that the Pauli corrections  $\prod_{j=1}^{ \lfloor \frac{v}{2}\rfloor }P_{(2j)}$ and the controlled-$P_{(j)}$ gates for even $j$ between the data and corresponding ancilla qubits can combine into zero-controlled gates using the circuit identities shown in Fig.~\ref{fig:CDR}.
  Figure~\ref{fig:Shor_SE} shows an example of measuring a stabilizer of weight 4.

 \begin{figure}[htbp]
 	\begin{algorithm}[H]
 		\caption{Modified Shor Syndrome Extraction} \label{alg:modified_shor}
 		
 		\begin{algorithmic}[1]
 			\State \textbf{Input:}  {A weight-$v$  CSS stabilizer
 			$g=(-1)^cP_{(1)}P_{(2)}\cdots P_{(v)}$,
 			where $c\in\{0,1\}$ and $P_{(j)}=P$ for all $j$ with $P\in\{X,Z\}$;}
 			a quantum state $\ket{\psi}$ encoded by a $2n$-qubit CE CSS code.

 			\State \textbf{Output:} Measurement outcome of $g$ on $\ket{\psi}$.

 			\State Prepare a $w$-CE cat state, {where $w=\lceil \frac{v}{2}\rceil$:}
 			$
 			\ket{\mathrm{cat}^{\mathrm{CE}}(w)} .
 			$
 			
 			\State Apply controlled-$P_{(j)}$ gates from the $j$-th ancilla qubit to the corresponding data qubit for all $j = 1$ to $v$.
 			
 			{
 			\State Apply Pauli corrections $\displaystyle \prod_{j=1}^{ \lfloor \frac{v}{2}\rfloor }P_{(2j)}$ to the data qubits.
 		}
 			
 			\State Measure the ancilla qubits in the $X$ basis and record the outcome $a \in \{0,1\}^{2w}$.
 			
 			\If{$\mathrm{wt}(a)$ is even}
 			\State \Return $0 + c \bmod 2$.
 			\Else
 			\State \Return $1 + c \bmod 2$.
 			\EndIf
 		\end{algorithmic}
 	\end{algorithm}
 \end{figure}

\bt \label{thm-9}
{A weight-$2w$ or $(2w-1)$ CSS stabilizer of a CE CSS code can be measured using Algorithm~\ref{alg:modified_shor} together with a $w$-CE cat state. The resulting syndrome extraction circuit uses only transversal two-qubit gates, single-qubit gates, and single-qubit measurements.}
\et

If $g$ is a stabilizer with phase $-1$, such as $-Z_1Z_2$, we perform the above procedure to measure $Z_1Z_2$ in practice and then compensate for the negative phase. Specifically, if the outcome $a$ has even weight, the measurement result for $g$ on $\ket{\psi}$ should be interpreted as $1$; if $a$ has odd weight, the result should be $0$.
\newline

{
We also provide a modified Steane syndrome extraction method for CE CSS codes in Appendix~\ref{app:steane}, which corrects the Steane-based construction proposed in~\cite{Ouy21} to satisfy the CE constraint.
}

\subsection{Extended Stabilizer Circuit Simulation for Coherent Errors}
\label{sec:extended_simulations}

The computationally efficient stabilizer tableau formalism~\cite{GK98}
cannot directly simulate CC errors, since coherent rotations are
non-Clifford operations.
To overcome this limitation, we develop an extended stabilizer simulation
procedure for CE CSS codes based on the identities in
Lemma~\ref{lemma:CE_identity}.

{
	In the standard stabilizer representation, an $N$-qubit Pauli operator
	$X^{\bfa}Z^{\bfb}$ is represented by two binary vectors
	$\bfa,\bfb\in\{0,1\}^N$.
	To capture coherent errors, we additionally represent each coherent
	rotation
	\[
	e^{i\theta Z_{(1)}\cdots Z_{(m)}}
	\]
	by a pair $(\theta,\bfc)$, where
	$\theta\in\mathbb{R}$ is the rotation angle and
	$\bfc\in\{0,1\}^N$ specifies its support.
	The simulation therefore tracks both the Pauli frame
	$(\bfa,\bfb)$ and a coherent error set $\mathcal{E}$ throughout the
	circuit.
}

As coherent rotations propagate through transversal two-qubit gates,
their supports are updated according to the identities in
Lemma~\ref{lemma:CE_identity}.
The remaining ingredient is to determine how coherent rotations are
updated during syndrome measurements.

{
	
	\bl \label{lemma:twirling_measurement}
	Let $g$ be the stabilizer measured by the modified Shor syndrome extraction circuit, and suppose the input state is
	$
	e^{i\theta Z^{\bfc}}E\ket{\phi},
	$
	where $\ket{\phi}$ is a code state, $E=X^{\bfa}Z^{\bfb}$ is the current Pauli frame, and
	$Z^{\bfc}$ is a coherent rotation.
	
	If $Z^{\bfc}$ commutes with $g$, then the coherent rotation does not affect the syndrome measurement and remains unchanged after the measurement.
	
	If $Z^{\bfc}$ anticommutes with $g$, then the syndrome measurement projects the state to
	\[
	\begin{cases}
		E\ket{\phi}, & \text{with probability }\cos^2\theta,\\[1ex]
		Z^{\bfc}E\ket{\phi}, & \text{with probability }\sin^2\theta,
	\end{cases}
	\]
	and the measured syndrome bit is flipped in the second case.
	
	\el
	
}

{
	Lemma~\ref{lemma:twirling_measurement} shows that every coherent
	rotation that anticommutes with the measured stabilizer is converted into
	an equivalent stochastic Pauli-$Z$ error with probability
	$\sin^2\theta$.
}

\begin{figure}[htbp]
	\begin{algorithm}[H]
		
		\caption{{Extended Stabilizer-Based Simulation of Circuit-Level Noise with Coherent Errors for CE CSS Codes}
		}
		\label{alg:ext_stabilizer}
		\begin{algorithmic}[1]
			\State \textbf{Input:} Circuit $\mathcal{C}$ with data and ancilla qubits (total $N$ qubits)
			\State Initialize Pauli error: $(\bfa, \bfb) \gets (\mathbf{0}^N, \mathbf{0}^N)$
			\State Initialize coherent error set: $\mathcal{E} \gets \emptyset$
			\For{each gate $G$ in $\mathcal{C}$}
			\If{$G$ is a {Pauli error}}
			\If{$G$ has an $X_j$ component}
			
			{
				\For{each $(\theta, \bfc) \in \mathcal{E}$}
				\If{$\bfc_j = 1$}
				\State Update: $\theta \gets - \theta$
				\EndIf
				\EndFor
			}
			
			\State Flip bit: $\bfa_j \gets \bfa_j \oplus 1$
			\EndIf
			\If{$G$ has a $Z_j$ component}
			\State Flip bit: $\bfb_j \gets \bfb_j \oplus 1$
			\EndIf
			\ElsIf{$G$ is a CC error of phase $\theta$}
			\For{each qubit $j$}
			\If{$\bfa_j = 1$}
			\State $\mathcal{E} \gets \mathcal{E} \cup \{(-2\theta, \bfe_j)\}$
			
			\EndIf
			\EndFor
			\State Merge coherent rotations with identical support
			\ElsIf{$G$ is $CX$, $C_0X$, $CZ$, or $C_0Z$}
			\State Update $(\bfa, \bfb)$ using standard stabilizer rules
			\For{each $(\theta, \bfc) \in \mathcal{E}$}
			\If{$\bfc_t = 1$}
			\If{$G$ is $CX(c, t)$ or $C_0X(c, t)$}
			\State Update: $\bfc_c \gets \bfc_c \oplus 1$
			
			\EndIf
			\If{{$G$ is $C_0X(c, t)$ }}
			\State Update: {$\theta \gets - \theta$}
			\EndIf
			\EndIf
			\EndFor
			\State Merge coherent rotations with identical  support

			\ElsIf{$G$ is an $M_X$ on ancilla $a$  following $CX_{a,d}$}
			
			\For{each $(\theta, \bfc) \in \mathcal{E}$}
			\If{ $\bfc_a=1$ or $\bfc_d=1$}
			\State Sample $r \sim \text{Uniform}(0, 1)$
			\If{$r < \sin^2(\theta)$}
			\State Flip: $\bfb \gets \bfb \oplus \bfc $
			\EndIf
			\State Remove $(\theta,\bfc)$ from $\mathcal{E}$
			
			\EndIf
			\EndFor
			\State Record outcome: $m_a \gets \bfb_a$
			\EndIf
			\EndFor
			\State \textbf{Output:} Final Pauli error vector $(\bfa, \bfb)$ and measurement outcomes $\{m_a\}$
		\end{algorithmic}
	\end{algorithm}
\end{figure}

We present the simulation procedure in Algorithm~\ref{alg:ext_stabilizer} for the FTEC circuits considered in this work,
 with additional details provided in Appendix~\ref{sec:sim_algorithms}.
{
	When a fault event is encountered, Lines~5--23 update the Pauli frame and coherent error set by incorporating the newly introduced error.
	When an ideal circuit operation or measurement is encountered, Lines~24--48 update the tracked errors according to the corresponding propagation and projection rules.
	Here, coherent rotations with identical support are merged according to $e^{i\theta Z^\bfc}e^{i\zeta Z^\bfc}=e^{i(\theta+\zeta)Z^\bfc}$.
}

{
	For the distance-three CE CSS codes considered in this paper, the
	algorithm exactly captures all dominant single- and double-fault events,
	while higher-order coherent interactions are simulated using the
	Pauli-twirling approximation.
}

{
Algorithm 2 is designed for the CE FTEC circuits considered in this work. The online circuit contains controlled-$X/Z$, zero-controlled-$X/Z$ gates, Pauli operations, and measurements. Although the algorithm can be extended to general Clifford circuits by including additional stabilizer propagation rules, such extensions require tracking more general coherent Pauli rotations and are beyond the scope of this work.
}

\subsection{Distance-3 Fault-Tolerant Error Correction}
 \begin{figure}[htbp]
\begin{algorithm}[H]
	\caption{Fault-Tolerant Error Correction for Distance-3 CE CSS Codes}
	\label{alg:control_flow}
	\begin{algorithmic}[1]
		
		\State \textbf{Input:} stabilizer generators of a CSS code, a syndrome lookup table, and an input quantum state
		\State \textbf{Output:} a Pauli correction
		
		\vspace{5pt}
		
		\State Measure $Z$-type stabilizers; record outcome as $m_Z^{(1)}$
		
		\If{$m_Z^{(1)}\neq 0$}
		\State Measure $Z$-type stabilizers, obtaining $m_Z^{(2)}$
		\State Apply the $X$-correction corresponding to $m_Z^{(2)}$
		\State Measure $X$-type stabilizers, obtaining $m_X^{(1)}$
		\If{$m_X^{(1)}\neq 0$}
		\State Measure $X$-type stabilizers, obtaining $m_X^{(2)}$
		\State Apply the $Z$-correction corresponding to $m_X^{(2)}$
		\EndIf

		\Else
		\State Measure $X$-type stabilizers, obtaining $m_X^{(1)}$
		\If{$m_X^{(1)}\neq 0$}
		\State Measure $Z$-type stabilizers, obtaining $m_Z^{(2)}$
		\If{$m_Z^{(2)}\neq 0$}
		\State Apply the $X$-correction corresponding to $m_Z^{(2)}$

		\State Measure $X$-type stabilizers, obtaining $m_X^{(2)}$
		\State Apply the $Z$-correction corresponding to $m_X^{(2)}$
		\Else
		\State Apply the $Z$-correction corresponding to $m_X^{(1)}$
		\EndIf
		\EndIf
		\EndIf
	\end{algorithmic}
\end{algorithm}
\end{figure}

\begin{figure*}[htbp]
	\centering
	\includegraphics[width=1\linewidth]{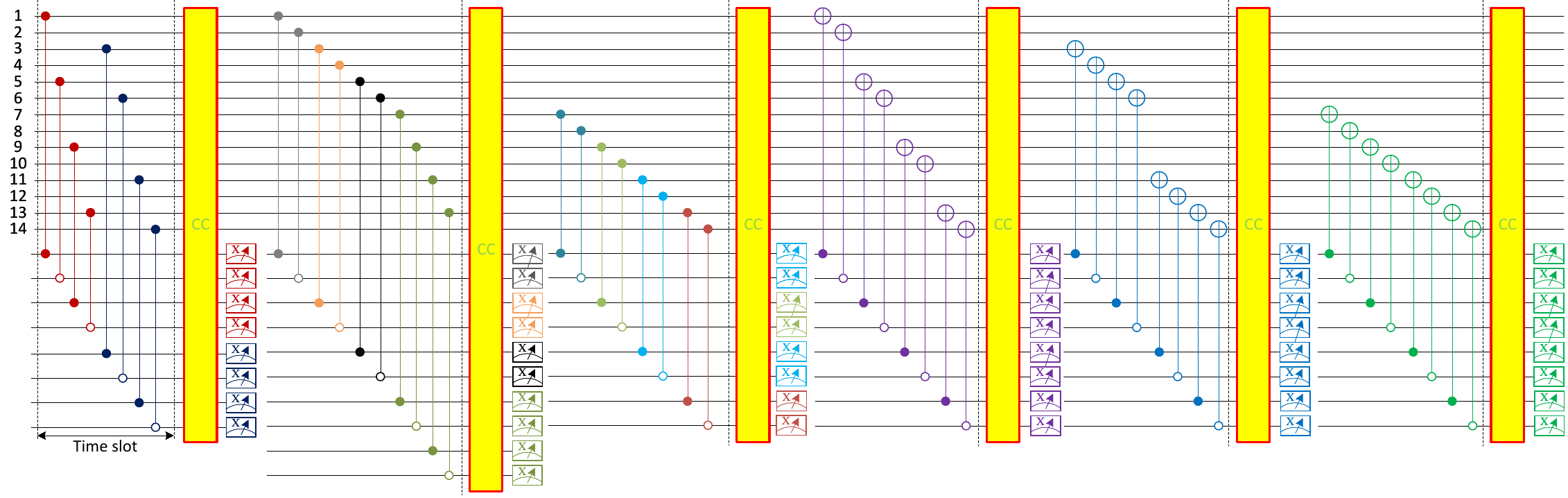}
\caption{Syndrome extraction circuit for the $[[14,1,3]]$ CE code using the modified Shor method. 
	Ancilla states are omitted. 
	Vertical boxes represent CC error layers inserted after each circuit depth, acting collectively on all qubits in the corresponding circuit layer, including idle data qubits.}
	\label{fig:1413_circuit}
\end{figure*}

{
Protocols with reduced syndrome measurement overhead have been studied in the literature, 
including conditional syndrome generation and optimized FTEC schemes~\cite{PVK97,DR20,TPB23,LL25}.
We propose a control flow for weak FTEC of distance-three CE CSS codes in Algorithm~\ref{alg:control_flow},
where at most two rounds of syndrome measurements are required for each type of stabilizer ($X$ or $Z$).
The proposed control flow exploits the CSS structure by conditionally processing $X$- and $Z$-type syndromes separately.
In particular, $X$-type errors are corrected before subsequent $Z$-type corrections, since residual $X$ errors may induce coherent rotations under the CC error model.

\bt \label{thm:weak_FTEC}
{The control flow in Algorithm~\ref{alg:control_flow}, combined with the modified Shor syndrome extraction circuit in Algorithm~\ref{alg:modified_shor}, implements a weak FTEC.}
\et
}

\begin{figure}[htbp]
	\centering
	\includegraphics[width=1\linewidth]{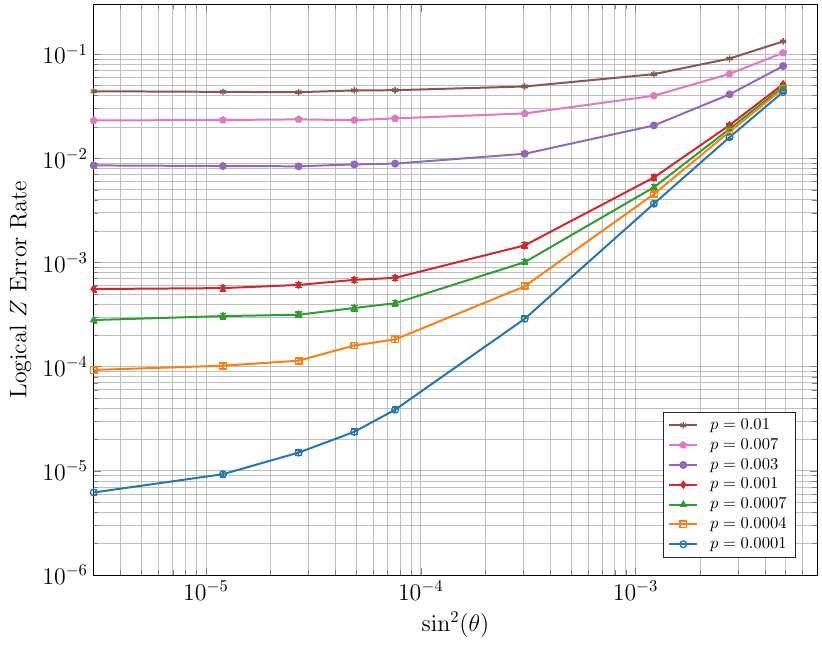}
	\caption{Logical $Z$ error rates of the $[[9,1,3]]$ rotated surface code under circuit-level noise with the Pauli-twirled coherent error model for various physical error rates $p$, where $\theta$ denotes the coherent rotation angle.} 	
	\label{fig:913_resultq}
\end{figure}

\begin{figure}[htbp]
	\centering
	\includegraphics[width=1\linewidth]{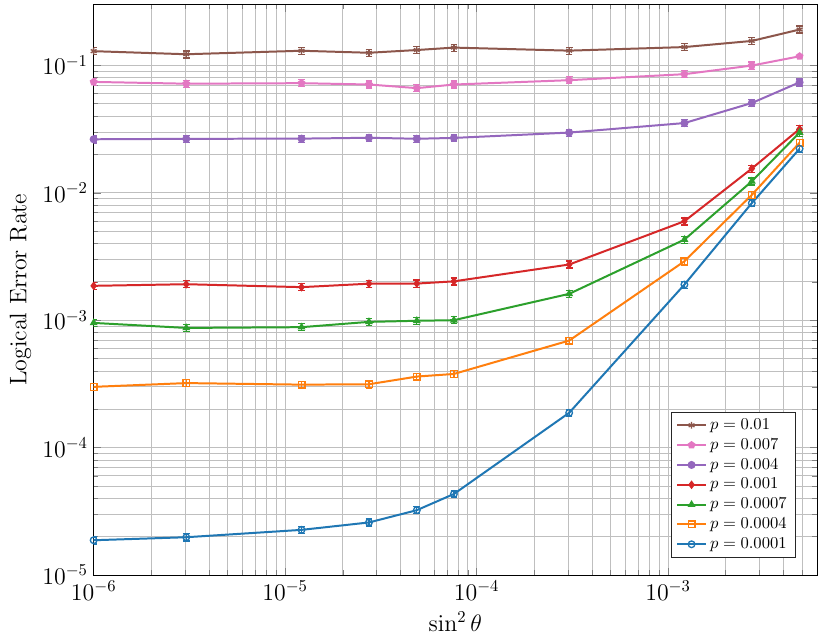}
		\caption{Logical error rates of the $[[7,1,3]]$ Steane code under circuit-level noise with the Pauli-twirled coherent error model for various physical error rates $p$, where $\theta$ denotes the coherent rotation angle.} 	
	\label{fig:713_resultq}
\end{figure}

\begin{figure}[htbp]
	\centering
	\includegraphics[width=1\linewidth]{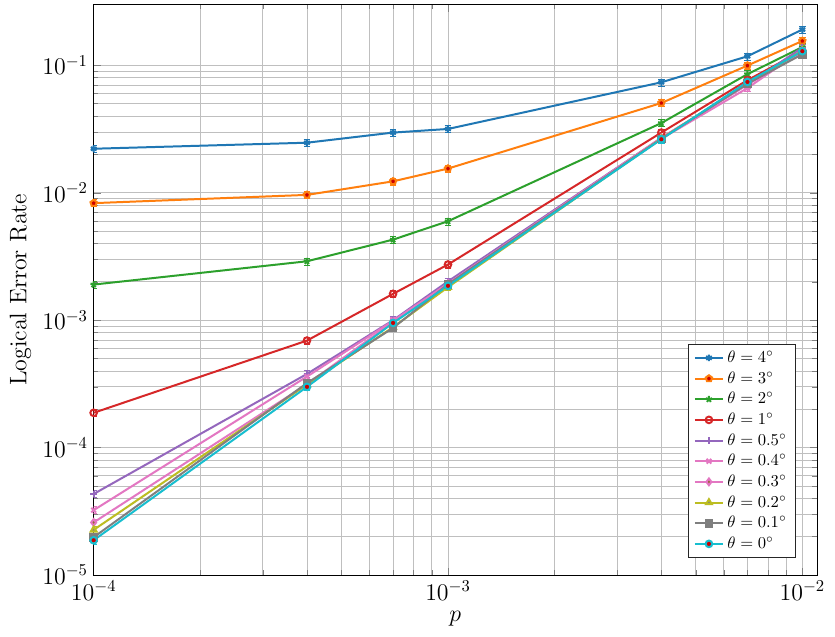}
	\caption{Logical error rates of the $[[7,1,3]]$ Steane code under circuit-level noise with the Pauli-twirled coherent error model for various $\theta$. }\label{fig:713_resultp}
\end{figure}

{
\section{Simulation}

We evaluate the proposed protocols under the circuit-level noise model with interleaved CC error layers. 
Specifically, after each circuit depth, a layer of collective coherent errors is applied to all qubits involved in the corresponding circuit layer, including idle data qubits. 
The syndrome extraction circuit for the $[[14,1,3]]$ CE CSS code is illustrated in Fig.~\ref{fig:1413_circuit}. 
The same layered CC noise model is also applied to the rotated surface code and the Steane code for comparison.

We first investigate the performance of conventional CSS codes under circuit-level noise with interleaved CC layers.
Unlike CE CSS codes, conventional CSS codes do not preserve their codespace under collective coherent $Z$ rotations.
As the rotation angle increases, the logical error rate becomes dominated by the effective Pauli-$Z$ error rate introduced by the Pauli-twirled coherent errors, and further degrades due to the accumulation of CC error layers throughout the syndrome extraction circuit.

We first consider the $[[9,1,3]]$ rotated surface code~\cite{BK98,BM07}.
We simulate three noisy rounds of syndrome extraction followed by a final perfect round, and the results are shown in Figure~\ref{fig:913_resultq}.
Here, the coherent rotation angle $\theta$ is typically small in practical settings.
For reference, $\theta=0.1^\circ,1^\circ,2^\circ,3^\circ,$ and $4^\circ$ correspond to
$\sin^2\theta \approx 3.0\times10^{-6}, 3.0\times10^{-4},
1.2\times10^{-3}, 2.7\times10^{-3},$ and $4.9\times10^{-3}$, respectively.
The details of the simulation procedure are provided in Appendix~\ref{sec:sim_algorithms}.
The results show that the effect of coherent errors becomes more pronounced as \(\theta\) increases relative to the physical error rate \(p\), leading to a significant increase in the logical error rate. This observation also supports the simplified noise model introduced in Section~\ref{sec:potential}, which captures the interplay between stochastic and coherent errors.

The same behavior is observed for the $[[7,1,3]]$ Steane code~\cite{Ste96a}.
Figures~\ref{fig:713_resultq} and~\ref{fig:713_resultp} show its performance under the control flow in Algorithm~\ref{alg:control_flow}.
}


{
Next, we simulate the performance of the $[[14,1,3]]$ CE Steane code under the control flow in Algorithm~\ref{alg:control_flow},
using the modified Shor syndrome extraction method and a lookup-table decoder.
The simulation is based on Algorithm~\ref{alg:ext_stabilizer}, which accurately captures single- and double-fault coherent error events as discussed in Appendix~\ref{sec:sim_algorithms}.
The circuit construction and simulation settings for the $[[14,1,3]]$ CE Steane code are provided in Appendix~\ref{app:1413}.

Figure~\ref{fig:1413_resultq} shows that the logical error rate of the $[[14,1,3]]$ CE Steane code remains nearly constant as the coherent rotation angle $\theta$ increases.
This demonstrates that the CE construction effectively suppresses the impact of CC errors in the considered circuit-level noise model.
}

 \begin{figure}[htbp]
 	\centering
 	\includegraphics[width=1\linewidth]{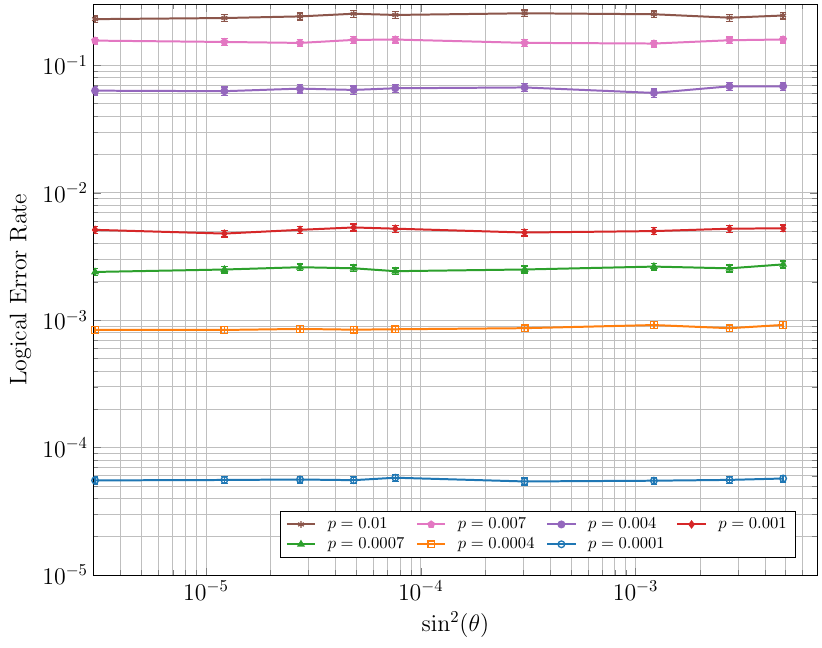}
 	\caption{Logical error rates of the $[[14,1,3]]$ CE Steane code under circuit-level noise with interleaved CC errors.
 	Compared with the \([[7,1,3]]\) code in Fig.~\ref{fig:713_resultp}, the dual-rail concatenated \([[14,1,3]]\) code exhibits a higher error floor, while its logical error rate becomes largely insensitive to the CC angle \(\theta\).
 	}
\label{fig:1413_resultq}
 \end{figure}

{
In the simulations, at least 1000 logical errors are collected for each data point.

The source files of our simulations are available at \cite{L26}.
}

\section{Conclusion} 
   
We have developed FTEC protocols for CE stabilizer codes under circuit-level coherent noise.
By constructing modified Shor- and Steane-type syndrome extraction circuits compatible with excitation-preserving constraints, we showed that CE codes can achieve reliable error correction while preserving their protection against coherent $Z$ rotations.
{In contrast to conventional CSS codes, where coherent errors accumulate across syndrome extraction cycles and degrade the logical error rate, the proposed CE constructions remain robust under the considered noise model.
}

{Our simulations focus on distance-3 CE CSS codes under the considered circuit-level noise model. Extending the analysis to larger-distance CE codes and more general coherent noise models remains an important direction for future work.
}

As future work, we plan to address the fault-tolerant preparation of $w$-CE cat states. Our approach will employ post-selected verification circuits~\cite{EG25}, where a failed verification triggers a re-preparation of the state. However, these circuits must be adapted to function effectively under the excitation-preserving constraints of $w$-CE cat states.

Another promising direction is the development of flagged syndrome extraction protocols~\cite{CR18a,CB18} in a parallel fashion~\cite{Rei20,LL23,LL25}, specifically tailored for CE codes.
The main challenges lie in designing flag circuits that are both excitation-preserving and fault-tolerant.

\section{Acknowledgements}\label{eq:acknow}

CYL acknowledges support from the National Science and Technology Council in Taiwan under Grants 113-2221-E-A49-114-MY3, 113-2119-M-A49-008-, and 114-2119-M-A49-006-, and support from the Ministry of Education, Taiwan, through the Taiwan Centers of Excellence (TCE) Program.

YO acknowledges support from EPSRC (Grant No. EP/W028115/1) and also the EPSRC funded QCI3 Hub under Grant No. EP/Z53318X/1.

\appendix

%


\newpage
\appendix

\section{Methods}

\subsection{CE CSS codes} \label{app:CE_codes}


We like to focus on CE codes that are not only stabilizer codes \cite{Got97}, but furthermore have a CSS structure \cite{CS96,Ste96a}. Stabilizer codes are described in the stabilizer formalism, which we briefly review.
Let $\mathcal{G}_n$ denote the $n$-qubit Pauli group, consisting of $n$-fold tensor products of Pauli matrices with an overall phase in $\{\pm 1, \pm i\}$.
Let $\bar{\mathcal{G}}_n=\{I,X,Y,Z\}^{\otimes n}$ denote the set of $n$-fold Pauli matrices with an overall phase $+1$.
The tensor product symbol is often omitted for simplicity. For example,
$X\otimes Y\otimes Z\otimes I\otimes I= XYZII = X_1 Y_2 Z_3$,
where the subscript indicates the qubit on which each Pauli operator acts nontrivially.
The weight of a Pauli operator is defined as the number of qubits on which it acts nontrivially.
We also define $X^{\bfa} Z^{\bfb}=\bigotimes_{i=1}^n X^{a_i}Z^{b_i} $ for $\bfa,\bfb\in\{0,1\}^n$.

A stabilizer code is defined by an Abelian subgroup $\mathcal{S} \subset \mathcal{G}_n$ that stabilizes the code space and $\cS$ does not contain the negative identity operator $-I$. This subgroup $\cS$ is referred to as a stabilizer group.  If $\mathcal{S}$ has $n - k$ independent generators, it defines an $[[n, k, d]]$ stabilizer code that encodes $k$ logical qubits into $n$ physical qubits. 
The parameter $d$ denotes the minimum distance of the code, which is the minimum weight of any Pauli error that has a nontrivial logical effect on the codespace.

The codespace of a stabilizer code is the joint $+1$ eigenspace of all the stabilizer generators. Measuring a set of stabilizer generators reveals information about any errors that may have occurred. Consequently, the resulting measurement outcomes are referred to as the error syndrome.

Each stabilizer generator is a Pauli operator with eigenvalues $\pm 1$. In practice, the measurement outcome is recorded as a binary value in ${0,1}$, corresponding to the eigenvalues $(-1)^0 = +1$ and $(-1)^1 = -1$, respectively. These binary outcomes comprise the syndrome bits, which are then used in the decoding process to identify and correct the underlying error.

One constructs CSS stabilizer code $\mathcal{C}(C_1, C_2)$~\cite{CS96,Ste96a} from two classical $n$-bit linear block codes $C_1$ and $C_2$, with dimensions $k_1$ and $k_2$, respectively, such that $C_2 \subset C_1$ and $k_2 < k_1$. The resulting CSS code uses $n$ physical qubits to encode $k = k_1 - k_2$ logical qubits.

We generalize the definition of CSS codes to include CE codes as follows.
The logical computational basis states of a CE CSS code  $\mathcal{C}(C_1, C_2)$ can be expressed as
\begin{align}
	\ket{j}_L = \frac{1}{\sqrt{|C_2|}} \sum_{\bfc \in C_2}   \ket{\bfy+\bfx^{(j)} + \bfc},
\end{align}
for each $j \in \{0,1\}^k$, where each $\bfx^{(j)} \in C_1$ is a representative of a distinct coset of $C_2$ in $C_1$. 
The vector $\bfy \in \{0,1\}^n$ allows for a global shift.
For conventional CSS codes,  $\bfy$ is trivial, i.e., $\bfy = \mathbf{0}$. 
These states form an orthonormal basis for the codespace of the CSS code.

If the classical codes $C_1$ and $C_2^\perp$ have minimum distances $d_1$ and $d_2^\perp$, respectively, then the minimum distance of the CSS code is
 $d= \min\{d_1, d_2^\perp\}$. 
Such a CSS code is denoted as an $[[n,k,d]]$ quantum code.

In the stabilizer formalism~\cite{Got97},
a CSS code $\mathcal{C}(C_1, C_2)$ has stabilizer group $\mathcal{S}$ that admits a generating set wherein each generator acts nontrivially with either only $X$ operators or only $Z$ operators on its support.
 Specifically, there is a bijection between the $X$-type stabilizer generators and the generator matrix of $C_2$, and between the $Z$-type stabilizer generators and the parity-check matrix of $C_1$. Namely, each $\bfg\in C_2$ defines a stabilizer $X^{\bfg}$ such that \begin{align}X^{\bfg} \ket{j}_L =\ket{j}_L,
 	\end{align} and for each $\bfh\in C_1^{\perp}$ it defines a stabilizer $(-1)^{\bfh \cdot \bfy} Z^{\bfh}$ such that 
 \begin{align}
 	 (-1)^{\bfh\cdot \bfy} Z^{\bfh} \ket{j}_L =\ket{j}_L,
 \end{align}
where $\cdot$ denotes the binary inner product.
 These $X$ and $Z$ stabilizers commute due to the nested structure $C_2 \subset C_1$.



As an example, the $[[4,1,2]]$ 2-CE CSS stabilizer code has stabilizer generators $XXXX$, $-ZZII$, and $-IIZZ$, with logical basis states 
\begin{align*} 
	\ket{0}_L &= \frac{1}{\sqrt{2}}\left(\ket{0110} + \ket{1001}\right),\\ \ket{1}_L &= \frac{1}{\sqrt{2}}\left(\ket{0101} + \ket{1010}\right). 
\end{align*}
 The corresponding logical operators are $\bar{X} = XXII$ and $\bar{Z} = IZZI$.
This is the smallest CE CSS stabilizer code capable of detecting any single-qubit Pauli error.

The the $[[4,1,2]]$ CE stabilizer code can also be constructed by concatenating the following two Bell states, each stabilized by $XX$, with the dual-rail code, according to Theorem~\ref{thm:construction}:
     	\begin{align*}
     			\ket{\beta_{01}}=&  \frac{1}{\sqrt{2}}\left(\ket{01}+\ket{10}\right),\\
     			\ket{\beta_{00}}=&\frac{1}{\sqrt{2}}\left(\ket{00}+\ket{11}\right).
     		\end{align*}

\subsection{Channel twirling}\label{app:twirling}
{
We analyze the effect of twirling the unitary CC  noise channel  $U(\rho)=U\rho U^\dag$,
where  $U = \exp(i \theta \sum_j Z_j)$.

For an arbitrary set of unitary operators $\mathcal{V}$ and a quantum channel $\mathcal{N}$, the $\mathcal{V}$-twirl of $\mathcal{N}$ is defined as~\cite{ESM+07}:
\[
\twirl(\mathcal{N}, \mathcal{V})(\rho) = \frac{1}{|\mathcal{V}|} \sum_{V \in \mathcal{V}} V^\dagger\, \mathcal{N}(V \rho V^\dagger)\, V.
\]
The Pauli expansion of $U$ only contains $I$- and $Z$-type Pauli operators
and $U= \frac{1}{2^n}\sum_{P\in\bar{cG}_n} \tr(P U) P  = \frac{1}{2^n}\sum_{P\in \{I,Z\}^{\otimes n} } \tr(PU) P$.
From Refs.~\cite{ESM+07,CCEE09,ouyang2014channel}, the Kraus operators of the
twirled channel $\twirl(\mathcal U,\bar{\mathcal G}_n)$ are given by
\begin{align}
	&\left\{
	2^{-n}P\,|\tr(PU)|:P\in\bar{\mathcal G}_n
	\right\}
	\nonumber\\
	=&
	\left\{
	2^{-n}P\,|\tr(PU)|:P\in\{I,Z\}^{\otimes n}
	\right\}
	\nonumber\\
	=&
	\left\{
	|\cos\theta|^{n-w(P)}
	|\sin\theta|^{w(P)}P:
	P\in\{I,Z\}^{\otimes n}
	\right\}
	\nonumber\\
	=&
	\left\{
	I|\cos\theta|,
	Z|\sin\theta|
	\right\}^{\otimes n},
\end{align}
where $w(P)$ denotes the number of $Z$ operators in $P$.
}

 {
 	Thus the Pauli-twirled coherent error channel is $\twirl(\mathcal U,\bar{\mathcal G}_n)= \mathcal{P}_{\mathrm{\textrm{coh}}}^{\otimes n}(\rho)$, where the single-qubit Pauli-twirled coherent error channel is $
 	 	\mathcal{P}_{\mathrm{\textrm{coh}}}(\sigma) =\cos^2\theta\sigma+\sin^2\theta Z\sigma Z.$
 The corresponding Pauli error probabilities are
 $
 (a_I,a_X,a_Y,a_Z)
 =
 (1-\sin^2\theta,0,0,\sin^2\theta).
 $
 
 The single-qubit depolarizing channel is
 \[
 \mathcal{D}_p(\rho)
 =
 (1-p)\rho+\frac{p}{3}
 (X\rho X+Y\rho Y+Z\rho Z),
 \]
 with Pauli probabilities
 $
 (b_I,b_X,b_Y,b_Z)
 =
 \left(1-p,\frac{p}{3},\frac{p}{3},\frac{p}{3}\right).
 $
 
 The composition of two Pauli channels is again a Pauli channel. 
 For
 \[
 \mathcal{P}_q=\mathcal{D}_p\circ\mathcal{P}_{\mathrm{\textrm{coh}}},
 \]
 the resulting probability of a Pauli error $R$ is obtained by summing all
 pairs $(P,Q)$ satisfying $PQ=R$ up to a phase:
 $
 q_R=\sum_{PQ=R}a_Pb_Q .
 $
 
 For the identity error, the contributing terms are
 $I\cdot I$ and $Z\cdot Z$:
 $
 q_I=a_Ib_I+a_Zb_Z
 =(1-\sin^2\theta)(1-p)+ \frac{p}{3} \sin^2\theta.
 $
%
 For the $X$ error, the contributing terms are
 $I\cdot X$ and $Z\cdot Y$:
 $
 q_X=a_Ib_X+a_Zb_Y
 =(1-\sin^2\theta)\frac{p}{3}+ \frac{p}{3} \sin^2\theta 
 =\frac{p}{3}.
 $
Similarly, for the $Y$ error,
 $
 q_Y=a_Ib_Y+a_Zb_X
 =(1-\sin^2\theta)\frac{p}{3}+\frac{p}{3} \sin^2\theta
 =\frac{p}{3}.
 $

 For the $Z$ error, the contributing terms are
 $I\cdot Z$ and $Z\cdot I$:
 $
 q_Z=a_Ib_Z+a_Zb_I
 =(1-\sin^2\theta)\frac{p}{3}+\sin^2\theta(1-p).
 $
Therefore, the final Pauli probabilities are
$
		q_I=1-p-\sin^2\theta+\frac{4}{3}p\sin^2\theta,$
		$q_X=q_Y=\frac{p}{3},$
		$q_Z=\sin^2\theta+\frac{p}{3}
		-\frac{4}{3}p\sin^2\theta .$
}

\subsection{Proof of Theorem \ref{thm:construction}}

	{
Since every codeword is stabilized by $-Z_{2j-1}Z_{2j}$, the state restricted to qubits $(2j-1,2j)$ must lie in the $+1$ eigenspace of $-Z_{2j-1}Z_{2j}$. The operator $-Z_{2j-1}Z_{2j}$ has eigenvalue $+1$ on $\ket{01}$ and $\ket{10}$, and eigenvalue $-1$ on $\ket{00}$ and $\ket{11}$. Hence every codeword has support only on ${\ket{01},\ket{10}}$ for each block $(2j-1,2j)$, implying that every block contains exactly one excitation. Since this holds for every $j=1,\ldots,n$, every codeword has exactly $n$ excitations. Therefore, the codespace is an $n$-CE subspace.
}
	
	It remains to verify that $\cS'$ defines a valid $[[2n,k,d'\geq d]]$ stabilizer code.
	
Note that {${X}^\CDR,{Z}^\CDR,$ and ${Y}^\CDR$} preserve that the commutation relations of $X,$ $Z,$ and $Y$. Hence
$\tau(g_i)$ and $\tau(g_j)$ commute since $g_i$ and $g_j$ commute by assumption.

Furthermore, each  $-Z_{2j-1}Z_{2j}$  commutes with $X_{2j-1}X_{2j}$ and $Y_{2j-1}X_{2j}$,
and therefore also commutes with $\tau(g_i)$ for all $i$.
Thus $\cS'$ is an Abelian group.

Since the stabilizers $-Z_{2j-1}Z_{2j}$ do not lie in the image of $\tau$ on $\mathcal{G}_n$, the elements $\tau(g_i)$ and $-Z_{2j-1}Z_{2j}$ are independent. As a result,   $\cS'$ defines a valid $[[2n,k]]$ stabilizer code.

Suppose that $\bar{X}_1,\dots, \bar{X}_k$, $\bar{Z}_1,\dots, \bar{Z}_k$ are $k$ logical operators 
that commute with $\cS$.
Let $\cN(\cS)$ denote the normalizer group of $\cS$ in $\cG_n$, i.e.,
$\cN(\cS)=\big\langle g_1,\dots,g_{n-k},$ $\bar{X}_1,\dots, \bar{X}_k$, $\bar{Z}_1,\dots, \bar{Z}_k \big\rangle \times
\{\pm1,\pm i\}.$
{
Then the normalizer of $\cS'$ is given by 
\begin{align*}
&\cN(\cS')\\
=&\big\langle \tau(g_1),\dots,\tau(g_{n-k}), \tau(\bar{X}_1),\dots,\tau(\bar{X}_k), \tau(\bar{Z}_1),\dots,\tau(\bar{Z}_k),\\
& -Z_1Z_2,\dots, -Z_{2n-1}Z_{2n}\big\rangle \times
\{\pm1,\pm i\}\\
=&\langle\tau(N(\cS)),-Z_1Z_2,\dots, -Z_{2n-1}Z_{2n}\big\rangle \times
\{\pm1,\pm i\}.
\end{align*}
}

{
Let $f$ be an element in $\tau(N(\cS))$. Then it can be verified that any element in 
$f\times  \langle -Z_1Z_2,\dots, -Z_{2n-1}Z_{2n}\big\rangle$ has weight no less than $f$.
This is because the restriction of $f$ to qubits $(2j-1,2j)$ is one of  $II, XX, YX, ZI$
due to the map $\tau$. Such components multiplied by $-Z_{2j-1}Z_{2j}$ become 
$ZZ, YY, XY, IZ$, up to global phases, and none of these replacements decreases the block weight.
Consequently, there always exists a minimum-weight element in $\cN(\cS')\setminus\cS'$ that lies in $\tau(\cN(\cS))$.
Therefore, it suffices to consider  minimum-weight elements in $\tau(\cN(\cS))\setminus \tau(\cS) $. 
}    

Since $\tau(g)$ has weight at least as large as $g$ for any $g\in\cG_n$, the minimum weight of any nontrivial logical operator in $\mathcal{N}(\mathcal{S}') \setminus \mathcal{S}'$ is at least $d$. Therefore, the minimum distance $d' \geq d$.


This CE construction can be seen as first extending the original stabilizer group using the mapping $\tau$ 
and then appending $n$ $Z$-type stabilizers. By Eq.~(\ref{eq:ZI}), $X$- or $Z$-type stabilizers remain $X$- or $Z$-type under $\tau$. Thus, if $\mathcal{S}$ defines a CSS code, then $\mathcal{S}'$ also defines a CSS code.

According to Eq.~(\ref{eq:UDR}), prior to encoding, an ancilla qubit  $\ket{0}$ at the $2j$-th location is stabilized by $Z_{2j}$. After encoding, this stabilizer is transformed into $-Z_{2j-1}Z_{2j}$. Therefore, the symplectic partner of $-Z_{2j-1}Z_{2j}$ is $X_{2j}$, which is encoded through the gate $C_0X_{2j}$ and remains unchanged as $X_{2j}$.

According to Eq.~(\ref{eq:UDR}), prior to encoding, an ancilla state $\ket{0}$ at the $2j$-th location is stabilized by $Z_{2j}$. Since $Z_{2j}$ and $X_{2j}$ are symplectic partners and are encoded to $-Z_{2j-1}Z_{2j}$ and $X_{2j}$, respectively, $X_{2j}$ remains the symplectic partner of $-Z_{2j-1}Z_{2j}$.

 \subsection{Proof of Lemma \ref{non-existence of 8,9 CE CSS}}

 Consider a CE CSS code  $\mathcal{C}(C_1, C_2)$ with basis states
 \begin{align*}
 	\ket{j}_L = \frac{1}{\sqrt{|C_2|}} \sum_{\bfc \in C_2}   \ket{\bfy+\bfx^{(j)} + \bfc},
 \end{align*}
 for each $j \in \{0,1\}^k$.

 For the CSS code to be a CE code, we require that each set
 \[
 W_j = \bfy + \bfx^{(j)} + C_2
 \]
 forms a constant-weight code, consisting only of codewords with Hamming weight \( w \), for some \( w \in \{1, \dots, n\} \). Let
 \[
 W = \bigcup_{j} W_j.
 \]
 Then \( W \) is a subset of \( C_1 \) with a shift \( \bfy \), and the minimum distance of \( W \) is at least the minimum distance of \( C_1 \). Consequently, the minimum distance of the resulting CE CSS code is lower bounded by the distance of \( C_1 \). In particular, to obtain a CE CSS code with distance at least 3, it is necessary that \( W \) also has minimum distance at least 3.

 Let $A(n, d, w)$ denote the maximum number of codewords in a constant-weight code of length $n$, weight $w$, and minimum distance $d$. It is known that $A(n, d, w) = A(n, d+1, w)$ for odd $d$ \cite{MS77}. In particular,  $A(n, 3, w) = A(n, 4, w)$. It suffices to consider constant-weight codes of weight 4.

 To maximize the distance of the dual code \( C_2^\perp \), we aim to maximize the number of codewords in \( C_2 \). The number of codewords in \( C_2 \) is equal to the number of codewords in \( W_j \), which is \( |W| / 2^{k_1 - k_2} \). Thus, we want \( W \) to have as many codewords as possible.
 
 Using known results on the upper bound for \( |W| \), which is given by \( A(n, d, w) \) from \cite{AVZ00}, we proceed as follows:

 For \( A(8, 4, w) \), the maximum is 14 for any feasible \( w \). This means that \( |C_2| \leq 7 \). However, since \( C_2 \) is a linear code, we must have \( |C_2| \leq 4 \). Thus, the dual code of \( C_2 \) has at least \( 2^8 / 4 = 64 \) codewords. From the table of linear codes, we know that the minimum distance is at most 2. Hence, it is not possible to have a length 8 CE CSS code.
 
 Similarly, for \( A(9, 4, w) \), the maximum is 18 for any feasible \( w \). This means that \( |C_2| \leq 9 \), but since \( C_2 \) is a linear code, \( |C_2| \leq 8 \). Therefore, the dual code of \( C_2 \) has at least \( 2^9 / 8 = 64 \) codewords. Again, from the table of linear codes, we know that the minimum distance is at most 2. Thus, it is not possible to have a length 9 CE CSS code.
 
 Using similar arguments we also cannot have a distance four CE CSS code with length $n\le 7$.

\subsection{Proof of Lemma \ref{non-existence-10CE}}

Assume that a $[[10,1,3]]$ CE CSS code can be constructed via the dual-rail concatenation method described in Theorem~\ref{thm:construction}, starting from a $[[5,1]]$ CSS code.

To ensure that the concatenated code has distance at least 3, the underlying $[[5,1]]$ CSS code must contain the all-$Z$ stabilizer $ZZZZZ$. This ensures that $Z$ errors affecting a single physical qubit on each block are detected. Since the total number of stabilizer generators is $n - k = 4$, this leaves at most three $X$-type stabilizers. These $X$-type stabilizers must have even weight to commute with $ZZZZZ$, so the $X$-stabilizer group corresponds to an even-weight classical $[5,3]$ code with minimum distance at least 2, denoted $C_2$.

To further ensure distance at least 3 for the concatenated code,  the dual code $C_2^\perp$ must have minimum distance at least 3. That is, $C_2^\perp$ must be a $[5,2,3]$ classical code.

To determine whether such a pair $(C_2, C_2^\perp)$ exists, we formulate an integer program based on their weight enumerators, subject to:
the MacWilliams identity between $C_2$ and $C_2^\perp$~\cite{MS77},
the minimum distance constraints (i.e., no codewords of weight less  than 3 in $C_2^\perp$), and the even-weight restriction on $C_2$.

An exhaustive search over all feasible solutions shows that no such pair exists.   Hence, it is impossible to construct a $[[10,1,3]]$ CE CSS code using Theorem~\ref{thm:construction}.

\subsection{Proof of Theorem \ref{thm-6-CE-12-14construction}} \label{app:constructions}

Starting from a $[[6,1,2]]$ CSS code with stabilizers
$X_1X_2$, $X_2X_3$, $X_4X_5$, $X_5X_6$, and
$Z_1Z_2Z_3Z_4Z_5Z_6$, we construct a $[[12,1,3]]$ CE CSS code via the dual-rail concatenation described in Theorem~\ref{thm:construction}.
The resulting $[[12,1,3]]$ CE CSS code is defined by the following stabilizers:
\begin{align*}
	g_1=&X_1X_2X_3X_4,&\\
	g_2=&X_3X_4X_5X_6,&\\
	g_3=&X_7X_8X_{9}X_{10},&\\
	g_4=&X_9X_{10}X_{11}X_{12},&\\
	g_5=&Z_1Z_3Z_5Z_7Z_9Z_{11},&\\
	g_6=&-Z_1Z_2,&\\
	g_7=&-Z_3Z_4,&\\
	g_8=&-Z_5Z_6,&\\
	g_9=&-Z_7Z_8.&\\
	g_{10}=&-Z_9Z_{10}.&\\
	g_{11}=&-Z_{11}Z_{12}.&
\end{align*}
The corresponding logical operators are  
\begin{equation*}
	\bar{X}=X_5X_6X_{11}X_{12}, \:\bar{Z}=Z_7Z_9Z_{12}.
\end{equation*}

Similarly, applying Theorem~\ref{thm:construction} to the $[[7,1,3]]$ Steane code~\cite{Ste96} yields a $[[14,1,3]]$ CE CSS code with the following stabilizers and logical operators:
{
	\begin{table}[H]
		\centering
	\begin{tabular}{ccccccccccccccc}
	$g_1 =$& $X$ & $X$ &$I$ &$I$ & $X$ & $X$ &$I$ &$I$ & $X$ & $X$ &$I$ &$I$ & $X$ & $X$ \\
	$g_2 =$& $I$ &$I$ & $X$ & $X$ & $X$ & $X$ &$I$ &$I$ &$I$ &$I$ & $X$ & $X$ & $X$ & $X$ \\
	$g_3 =$& $I$ &$I$ &$I$ &$I$ &$I$ &$I$ & $X$ & $X$ & $X$ & $X$ & $X$ & $X$ & $X$ & $X$ \\
	$g_4 =$& $Z$ &$I$ &$I$ &$I$ &$Z$ &$I$ &$I$ &$I$ &$Z$ &$I$ &$I$ &$I$ &$Z$ &$I$ \\
	$g_5=$ &$I$&$I$&$Z$&$I$&$Z$&$I$&$I$&$I$&$I$&$I$&$Z$&$I$&$Z$&$I$ \\
	$g_6=$ &$I$&$I$&$I$&$I$&$I$&$I$&$Z$&$I$&$Z$&$I$&$Z$&$I$&$Z$&$I$ \\
	$g_7 =$&$-Z$ &$Z$ &$I$ &$I$ &$I$ &$I$ &$I$ &$I$ &$I$ &$I$ &$I$ &$I$ &$I$ &$I$ \\
	$g_8 =$& $-I$ &$I$ &$Z$ &$Z$ &$I$ &$I$ &$I$ &$I$ &$I$ &$I$ &$I$ &$I$ &$I$ &$I$ \\
	$g_9 =$& $-I$ &$I$ &$I$ &$I$ &$Z$ &$Z$ &$I$ &$I$ &$I$ &$I$ &$I$ &$I$ &$I$ &$I$ \\
	$g_{10} =$& $-I$ &$I$ &$I$ &$I$ &$I$ &$I$ &$Z$ &$Z$ &$I$ &$I$ &$I$ &$I$ &$I$ &$I$ \\
	$g_{11} =$& $-I$ &$I$ &$I$ &$I$ &$I$ &$I$ &$I$ &$I$ &$Z$ &$Z$ &$I$ &$I$ &$I$ &$I$ \\
	$g_{12}=$& $-I$ &$I$ &$I$ &$I$ &$I$ &$I$ &$I$ &$I$ &$I$ &$I$ &$Z$ &$Z$ &$I$ &$I$ \\
	$g_{13}=$& $-I$ &$I$ &$I$ &$I$ &$I$ &$I$ &$I$ &$I$ &$I$ &$I$ &$I$ &$I$ &$Z$ &$Z$\\
	$\bar{X}=$ &$X$&$X$&$X$&$X$&$X$&$X$&$X$&$X$&$X$&$X$&$X$&$X$&$X$&$X$  \\
	$\bar{Z}=$ &$Z$&$I$&$Z$&$I$&$Z$&$I$&$Z$&$I$&$Z$&$I$&$Z$&$I$&$Z$&$I$.
\end{tabular}
\end{table}
}

{
	Starting from the $X$-type stabilizers of the $[[7,1,3]]$ Steane code~\cite{Ste96} together with the logical operator
	\[
	\bar{Z}=Z_1Z_2Z_3Z_4Z_5Z_6Z_7,
	\]
	 we construct a $[[14,3,3]]$ CE CSS code via the dual-rail concatenation described in Theorem~\ref{thm:construction}.
The resulting code is defined by the following stabilizers:
}
\begin{table}[H]
	\centering
	\begin{tabular}{ccccccccccccccc}
		$g_1=$ &$X$&$X$&$I$&$I$&$X$&$X$&$I$&$I$&$X$&$X$&$I$&$I$&$X$&$X$  \\
		$g_2=$ &$I$&$I$&$X$&$X$&$X$&$X$&$I$&$I$&$I$&$I$&$X$&$X$&$X$&$X$ \\
		$g_3=$ &$I$&$I$&$I$&$I$&$I$&$I$&$X$&$X$&$X$&$X$&$X$&$X$&$X$&$X$ \\
		$g_4=$ &$Z$&$I$&$Z$&$I$&$Z$&$I$&$Z$&$I$&$Z$&$I$&$Z$&$I$&$Z$&$I$  \\
		$g_5=$ &$-Z$&$Z$&$I$&$I$&$I$&$I$&$I$&$I$&$I$&$I$&$I$&$I$&$I$&$I$ \\
		$g_6=$ &$-I$&$I$&$Z$&$Z$&$I$&$I$&$I$&$I$&$I$&$I$&$I$&$I$&$I$&$I$ \\
		$g_7=$ &$-I$&$I$&$I$&$I$&$Z$&$Z$&$I$&$I$&$I$&$I$&$I$&$I$&$I$&$I$ \\
		$g_{8}=$ &$-I$&$I$&$I$&$I$&$I$&$I$&$Z$&$Z$&$I$&$I$&$I$&$I$&$I$&$I$ \\
		$g_{9}=$ &$-I$&$I$&$I$&$I$&$I$&$I$&$I$&$I$&$Z$&$Z$&$I$&$I$&$I$&$I$ \\
		$g_{10}=$ &$-I$&$I$&$I$&$I$&$I$&$I$&$I$&$I$&$I$&$I$&$Z$&$Z$&$I$&$I$ \\
		$g_{11}=$ &$-I$&$I$&$I$&$I$&$I$&$I$&$I$&$I$&$I$&$I$&$I$&$I$&$Z$&$Z$ \\
	\end{tabular}
\end{table}

\subsection{Proof of Lemma \ref{prop:CNOT}}

{
Consider a CE code with stabilizer group $\mathcal{S}'$ generated by $\tau(g_1), \dots, \tau(g_{n-k})$ and $-Z_1Z_2, \dots, -Z_{2n-1}Z_{2n}$,
where each $\tau(g_j)$ is a stabilizer of either $X$-type or $Z$-type with phase $+1$.

Suppose we have two code blocks, and a transversal CNOT gate is applied, using the first block as control and the second as target. 

Let $g^X$ and $g^Z$ denote arbitrary $X$-type and $Z$-type Pauli operators, respectively.
\begin{align*}
	g^X\otimes I  \mapsto  g^X\otimes g^X\\
	I\otimes g^X   \mapsto  I \otimes g^X\\
	g^Z\otimes I  \mapsto  g^Z\otimes I\\
	I\otimes g^Z   \mapsto  g^Z \otimes g^Z
\end{align*}
Therefore, every logical $X$ operator and every logical $Z$ operator propagate according to the standard CNOT conjugation rules. Hence the transversal CNOT realizes the logical operator $\mathrm{CNOT}^{\otimes k}$ on the encoded qubits.

Moreover,
\[
I,X,Y,Z\xmapsto{\tau}II,XX,YX,ZI,
\]
whose second component is always either $I$ or $X$.
Hence the correction operator
$\prod_{j=1}^n X_{2j}$
commutes with every operator
$\tau(g)$, $g\in\mathcal G_n$.

Likewise, every stabilizer of the form $\tau(g)$ propagates according to the standard CNOT conjugation rules and commutes with the correction operator $\prod_{j=1}^n X_{2j}$.
}

Next, we analyze the action of this transversal CNOT gate followed by  correction $\prod_{j=1}^n X_{2j}$ on the weight-2 stabilizers $-Z_{2i-1}Z_{2i}$.

The relevant stabilizers before applying the transversal CNOT are:
\begin{align*}
	-Z_1Z_2&\otimes I^{\otimes 2n}\\
	\vdots&\\
	-Z_{2n-1}Z_{2n}&\otimes I^{\otimes 2n}\\
	-  I^{\otimes 2n}&\otimes Z_1Z_2 \\
	\vdots&\\
	-    I^{\otimes 2n}&\otimes Z_{2n-1}Z_{2n}\\
\end{align*}
After applying the transversal CNOT gate (control: first block; target: second block), these stabilizers are transformed into:
\begin{align*}
	-Z_1Z_2&\otimes I^{\otimes 2n}\\
	\vdots&\\
	-Z_{2n-1}Z_{2n}&\otimes I^{\otimes 2n}\\
	-  Z_1Z_2&\otimes Z_1Z_2 \\
	\vdots&\\
	-   Z_{2n-1}Z_{2n}&\otimes Z_{2n-1}Z_{2n}\\
\end{align*}
Now, 
{ multiplying each stabilizer  $-Z_{2i-1}Z_{2i} \otimes Z_{2i-1}Z_{2i}$  with the corresponding  $-Z_{2i-1}Z_{2i} \otimes I^{\otimes 2n}$ yields:}
\begin{align*}
	-Z_1Z_2&\otimes I^{\otimes 2n}\\
	\vdots&\\
	-Z_{2n-1}Z_{2n}&\otimes I^{\otimes 2n}\\
	I^{\otimes 2n}&\otimes Z_1Z_2 \\
	\vdots&\\
	I^{\otimes 2n}&\otimes Z_{2n-1}Z_{2n}.
\end{align*}

The second block's stabilizers now have phase $+1$ instead of the original $-1$, effectively flipping their signs.
To restore the code space, we apply the Pauli correction $\prod_{j=1}^n X_{2j}$ to the second block (target). Therefore, the correction restores the stabilizer generators while leaving the logical action unchanged. Hence the overall circuit implements the logical transversal $\mathrm{CNOT}^{\otimes k}$.

The   correction  $X_2X_4\cdots X_{2n}$ can be incorporated into the transversal CNOT gates by using zero-controlled NOT gates by the circuit identity in Fig~\ref{fig:CDR}.

\subsection{Proof of Theorem \ref{thm-9}}

{
The modified Shor extraction differs from the conventional Shor extraction only by fixed single-qubit Pauli operations used for preparing the CE cat state and compensating the controlled-Pauli operations. 
}

Let $\ket{\psi}$ be a quantum state to be measured.
{Suppose we aim to measure a weight-$v$ stabilizer
$g=(-1)^c P_{(1)} P_{(2)} \cdots P_{(v)}$  on  $\ket{\psi}$,
	where  $c \in \{0,1\}$ and $P_{(j)}=P$ for all $j$, with $P\in\{X,Z\}$.}
	
{
Let $\displaystyle P_{\textrm{even}}=\prod_{j=1}^{ \lfloor \frac{v}{2}\rfloor }P_{(2j)}$ be the Pauli correction and
$\displaystyle P_{\textrm{odd}}=\prod_{j=1}^{ \lceil \frac{v}{2}\rceil }P_{(2j-1)}$.
Then  
 $(-1)^cg= P_{\textrm{odd}}P_{\textrm{even}}. $
}

{For $v=2w$ or $2w-1$, a $w$-CE cat state $\ket{\mathrm{cat}^{\mathrm{CE}}(w)} = \frac{1}{\sqrt{2}}(\ket{01}^{\otimes w} + \ket{10}^{\otimes w})$ is prepared.
}

  	 
\noindent 1)
Apply controlled-$P_{(j)}$ gates with the ancilla as the control qubit and the corresponding data qubits as the target qubits:
 \begin{align*}
 	&\ket{\psi}\otimes \frac{1}{\sqrt{2}}\left(  \ket{01}^{\otimes w}+ \ket{10}^{\otimes w} \right)\\
	&\xrightarrow{ CP_{(j)}} 	
	{\frac{1}{\sqrt{2}}\left(  \left(
	 P_{\textrm{even}} \ket{\psi}\right) \ket{01}^{\otimes w} 
	  +  \left(P_{\textrm{odd}} \ket{\psi}\right) \ket{10}^{\otimes w} \right)}\\
	=&  {\left(
	P_{\textrm{even}} \otimes I^{\otimes 2w}\right) \frac{1}{\sqrt{2}}\left(\ket{\psi}   \ket{01}^{\otimes w} 
	+  (-1)^c \left(   g\ket{\psi}\right) \ket{10}^{\otimes w} \right).}
\end{align*}

\noindent 2) Apply  correction  {$P_{\textrm{even}}=\displaystyle \prod_{j=1}^{ \lfloor \frac{v}{2}\rfloor }P_{(2j)}$} to the data qubits:

{
\begin{align*}
		\xrightarrow{P_{\textrm{even}}} 	& \frac{1}{\sqrt{2}}\left(\ket{\psi}   (I {X}\ket{00})^{\otimes w} \right.
		+   (-1)^c \left. \left(   g\ket{\psi}\right) (I {X}\ket{11})^{\otimes w} \right)
\end{align*}
}
\noindent 3) Measuring the ancilla qubits in the $X$ basis is equivalent to applying Hadamard gates followed by measurements in the $Z$ basis.

	Apply Hadamard gates to each ancilla qubit:

\begin{align}
	\xrightarrow{H_j} 	& \frac{1}{\sqrt{2}}\left(\ket{\psi}   (IZ\ket{++})^{\otimes w} \right.
		+   (-1)^c\left. \left(   g\ket{\psi}\right) (IZ\ket{--})^{\otimes w} \right) \notag\\
		=	&  \frac{\left(I^{\otimes 2n}\otimes (IZ)^{\otimes w}\right)}{\sqrt{2}}\left(\ket{\psi}   \ket{++}^{\otimes w} \right.+ (-1)^c  \left. \left(   g\ket{\psi}\right) \ket{--}^{\otimes w} \right) \notag\\
		=	& \frac{\left(I^{\otimes 2n}\otimes (IZ)^{\otimes w}\right)}{\sqrt{2} 2^{w-1}} \left(\frac{I+ (-1)^c g}{2}\ket{\psi}  \left( \sum_{v\in\{0,1\}^{2w}: \atop \textrm{wt}(v) \textrm{ even}}  \ket{v} \right) \right. \notag\\
		&+ \left. \frac{I- (-1)^c g}{2}\ket{\psi}  \left( \sum_{v\in\{0,1\}^{2w}: \atop \textrm{wt}(v)=1\mod 2}  \ket{v} \right) \right). \label{eq:modified_shor}
 \end{align}
 Then  measure the ancilla qubits in the $Z$ basis. 
 Note that the operator $\left(I^{\otimes 2n}\otimes (IZ)^{\otimes w}\right)$ does not affect the measurement   in the $Z$ basis.
 Let the outcome be $a \in \{0,1\}^{2w}$.  According to Eq.~(\ref{eq:modified_shor}), if $\mathrm{wt}(a)$ is even, the reported syndrome bit is shifted by c modulo 2. if $\mathrm{wt}(a)$ is odd,  the reported syndrome bit  is $1 + c \bmod 2$.

{Therefore, the measurement statistics are identical to those of conventional Shor syndrome extraction, except that the stabilizer phase $(-1)^c$ introduces a deterministic offset in the reported syndrome.}

\subsection{Modified Steane Syndrome Extraction}\label{app:steane}

\begin{figure}[htbp]
	\includegraphics[width=0.99\linewidth]{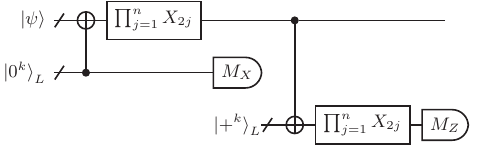}
	\caption{Modified Steane syndrome extraction. The gate $ \prod_{j=1}^n X_{2j}$ denotes a Pauli correction associated with the logical transversal CNOT operation. }\label{fig:Steane_SE}
\end{figure}

Steane introduced a syndrome extraction technique for CSS codes~\cite{Ste97L}. In Fig.~\ref{fig:Steane_SE}, we present a modified version of this method, in which two additional Pauli corrections are inserted to complement the action of the transversal CNOT gates, as justified by Lemma~\ref{prop:CNOT}. 
{By Lemma~\ref{prop:CNOT}, the modified transversal CNOT gate together with the Pauli correction $\prod_{j=1}^nX_{2j}$ implements a logical transversal CNOT gate on CE CSS codes.}
The states $\ket{0^k}_L$ and $\ket{+^k}_L$ denote the logical basis states of an $[[n,k]]$ CSS code, stabilized by all logical $Z$ and logical $X$ operators, respectively. {In a CE CSS code, these logical states are immune to coherent errors and are therefore well suited for fault-tolerant syndrome extraction.}

We can incorporate the correction $\prod_{j=1}^n X_{2j}$ into the transversal CNOT gates by replacing them with zero-controlled NOT gates. 
Alternatively, we can omit the second application of $\prod_{j=1}^n X_{2j}$ by flipping the corresponding measurement outcomes.

\bt \label{thm-10}
The circuit shown in Fig.~\ref{fig:Steane_SE} implements a  modified Steane syndrome extraction procedure for CE CSS codes. 

\et

We remark that the ancilla states required for the modified Steane syndrome extraction can be prepared offline via state distillation~\cite{LZB17,ZLB18}.

\begin{proof}

To extract the syndrome of a quantum state $\ket{\psi}$ encoded by an $[[n,k]]$ CE CSS code, we employ the logical ancilla states $\ket{0_L^k}$ and $\ket{+_L^k}$ encoded by the same code. By construction, these logical states are immune to coherent errors. {By Lemma~\ref{prop:CNOT}, the modified transversal CNOT gate together with the Pauli correction $\prod_{j=1}^nX_{2j}$ implements a logical transversal CNOT gate on CE CSS codes. The Pauli correction restores the original $-1$ phases of the weight-2 $Z$-type stabilizers.}
 {Consequently, the circuit in Fig.~\ref{fig:Steane_SE} performs the same logical syndrome extraction as the conventional Steane syndrome extraction circuit, with an additional fixed Pauli correction required for CE CSS codes.
}

	The circuit in Fig.~\ref{fig:Steane_SE} uses two transversal CNOT gates and two bitwise measurements, $M_X$ and $M_Z$, in the $X$ and $Z$ bases, respectively. The first CNOT gate propagates any $Z$ error on $\ket{\psi}$ to the $\ket{0^k}_L$ ancilla, while the second CNOT propagates $X$ errors to the $\ket{+^k}_L$ ancilla. We choose the ancilla codewords so that these operations do not affect the ancilla's logical state.
	

	Let the measurement outcomes from $M_X$ and $M_Z$ be $m^X_1, \dots, m^X_n$ and $m^Z_1, \dots, m^Z_n$, respectively. 
	
	For an $X$-type stabilizer with support vector $(a_1, \dots, a_n) \in \{0,1\}^n$ and phase $+1$, the measurement outcome is given by
	\begin{align*}
		\sum_{i=1}^n   a_i \cdot m^X_i   \mod 2.
	\end{align*}
	For a $Z$-type stabilizer with support vector $(b_1, \dots, b_n) \in \{0,1\}^n$ and phase $+1$, the outcome is similarly
	\begin{align*}
		\sum_{i=1}^n   b_i \cdot m^Z_i   \mod 2.
	\end{align*}
If the $Z$-type stabilizer instead has phase $-1$, the correct outcome must compensate for the phase, and is given by
	\begin{align*}
		1+ \sum_{i=1}^n   b_i \cdot m^Z_i   \mod 2.
	\end{align*}
This procedure ensures that all syndrome bits corresponding to the $X$- and $Z$-type stabilizers are correctly extracted, including cases with $-1$ phase factors.


{
	The circuit consists of transversal CNOT gates and fixed Pauli corrections, where the latter are single-qubit operations. 
}

\end{proof}
The modified Steane-type extraction requires specific ancillary state preparations to implement Clifford and non-Clifford operations~\cite{BK05,LZB17,ZLB18}. In our approach, these states are prepared through magic state distillation. Integrating magic state distillation into the CE framework enables universal gate implementation without compromising fault tolerance.

\subsection{Fault-tolerant syndrome extraction example}

\begin{example}
	Consider the $[[4,1,2]]$ CE CSS code with logical state $\ket{+}_L = \frac{1}{2} \left( \ket{0110} + \ket{1001} + \ket{0101} + \ket{1010} \right)$ and a correction $IXIX$. Suppose $\ket{\psi} = \ket{+}_L$ is measured using the modified Steane syndrome extraction. In the second part of the circuit, we have:
	\begin{align*}
		&\ket{+}_L\otimes \ket{+}_L\\
		\xrightarrow{CX}&\ket{+}_L\otimes \frac{1}{2}\left(\ket{0000}+\ket{1111}+\ket{0011}+\ket{1100}\right)\\
		\xrightarrow{IXIX}&\ket{+}_L\otimes \ket{+}_L.
	\end{align*}
	Measuring the second codeword in the $Z$ basis then yields one of the outcomes $0110$, $1001$, $0101$, or $1010$, each with equal probability. For instance, if the outcome is $0101$, we infer that measuring the $-Z_1 Z_2$ stabilizer gives the result $(-1) \cdot (-1)^0 \cdot (-1)^1 = +1$, and similarly for $-Z_3 Z_4$.
	
	If the second correction $IXIX$  is omitted, the $Z$-basis measurement yields one of $0000$, $1111$, $0011$, or $1100$, each with equal probability. Since the support vector of $IXIX$ is $0101$, we can adjust the measurement outcomes by bitwise addition modulo 2 with this support vector, obtaining the corrected outcomes $0101$, $1010$, $1001$, and $0110$, which match those from the version with the correction applied.

	\hfill~$\square$
\end{example}

{
	\subsection{Simulation algorithms for coherent errors}\label{sec:sim_algorithms}
	
	Simulating circuit-level noise in the presence of layers of CC errors is generally challenging.
	Since coherent rotations are non-Clifford operations, they cannot be efficiently tracked within the stabilizer formalism.
	Moreover, the number of coherent rotations may grow beyond the number of qubits due to gate propagation, rendering an exact simulation computationally intractable.

	\subsubsection{General CSS codes with the Pauli-twirling approximation}
	
	In this paper, we employ the Pauli-twirling approximation to simulate coherent errors in surface codes and the Steane code.
	Specifically, each coherent rotation $e^{i\theta Z_j}$ is replaced by a Pauli-$Z$ dephasing channel with error probability $\sin^2\theta$.
	Consequently, each layer of CC errors is modeled as independent $Z$ errors acting on all data and ancilla qubits.
	
	For surface codes, it has been shown that, under a single layer of CC errors and perfect syndrome extraction, the Pauli-twirling approximation underestimates the logical error rate when the rotation angle $\theta$ is below a certain threshold~\cite{BEK+18}.

	
	\begin{figure}[htbp]   
		\begin{center}
			
			\begin{tikzpicture}[scale=1.0]
				
				\draw[gray!20, thin] (0,0) grid (4,4);
				
				\coordinate (D1) at (2,4);
				\coordinate (D2) at (3,3);
				\coordinate (D3) at (4,2);
				\coordinate (D4) at (1,3);
				\coordinate (D5) at (2,2);
				\coordinate (D6) at (3,1);
				\coordinate (D7) at (0,2);
				\coordinate (D8) at (1,1);
				\coordinate (D9) at (2,0);
				
				\coordinate (A1) at (1,4); 
				\coordinate (A2) at (0,1); 
				\coordinate (A3) at (4,3); 
				\coordinate (A4) at (3,0); 
				\coordinate (A5) at (2,3); 
				\coordinate (A6) at (1,2); 
				\coordinate (A7) at (3,2); 
				\coordinate (A8) at (2,1); 
				
				\fill[xcol] (D1) -- (D2) -- (D5) -- (D4) -- cycle; 
				\fill[xcol] (D5) -- (D6) -- (D9) -- (D8) -- cycle; 
				\fill[xcol] (D2) -- (A3) -- (D3) -- cycle;       
				\fill[xcol] (D7) -- (A2) -- (D8) -- cycle;       
				
				\fill[zcol] (D4) -- (D5) -- (D8) -- (D7) -- cycle; 
				\fill[zcol] (D2) -- (D3) -- (D6) -- (D5) -- cycle; 
				\fill[zcol] (D1) -- (A1) -- (D4) -- cycle;       
				\fill[zcol] (D6) -- (A4) -- (D9) -- cycle;       
				
				\foreach \x in {0,1,2,3,4} {
					\foreach \y in {0,1,2,3,4} {
						\node[circle, draw, gray!30, fill=white, inner sep=0pt, minimum size=0.6cm] at (\x,\y) {};
					}
				}
				
				\foreach \i in {1,...,9} {
					\node[circle, draw, thick, fill=white, inner sep=1pt, minimum size=0.6cm, font=\small] at (D\i) {$D_{\i}$};
				}
				
				\node[circle, draw, fill=blue!10, inner sep=1pt, minimum size=0.6cm, font=\small] at (A1) {$A_{1}$};
				\node[circle, draw, fill=red!10, inner sep=1pt, minimum size=0.6cm, font=\small] at (A2) {$A_{2}$};
				\node[circle, draw, fill=red!10, inner sep=1pt, minimum size=0.6cm, font=\small] at (A3) {$A_{3}$};
				\node[circle, draw, fill=blue!10, inner sep=1pt, minimum size=0.6cm, font=\small] at (A4) {$A_{4}$};
				\node[circle, draw, fill=red!10, inner sep=1pt, minimum size=0.6cm, font=\small] at (A5) {$A_{5}$};
				\node[circle, draw, fill=blue!10, inner sep=1pt, minimum size=0.6cm, font=\small] at (A6) {$A_{6}$};
				\node[circle, draw, fill=blue!10, inner sep=1pt, minimum size=0.6cm, font=\small] at (A7) {$A_{7}$};
				\node[circle, draw, fill=red!10, inner sep=1pt, minimum size=0.6cm, font=\small] at (A8) {$A_{8}$};

			\end{tikzpicture}
		\end{center}
		\caption{Two-dimensional layout of the rotated surface code with distance $d=3$.
			Data qubits are labeled $D_1$--$D_9$, and ancilla qubits are labeled $A_1$--$A_8$.
			Green plaquettes denote $Z$ stabilizers, while red plaquettes denote $X$ stabilizers.}
		\label{fig:surface_d3}
	\end{figure}
	
	Figure~\ref{fig:surface_d3} illustrates the two-dimensional layout of the $[[9,1,3]]$ rotated surface code, which is defined by the following stabilizer generators:
	\begin{equation}
		\begin{aligned}
			S^X_{1} &= X_1 X_2 X_4 X_5, & S^X_{2} &= X_7 X_8, \\
			S^X_{3} &= X_2 X_3, & S^X_{4} &= X_5 X_6 X_8 X_9, \\
			S^Z_{1} &= Z_1 Z_4, & S^Z_{2} &= Z_4 Z_5 Z_7 Z_8, \\
			S^Z_{3} &= Z_2 Z_3 Z_5 Z_6, & S^Z_{4} &= Z_6 Z_9,
		\end{aligned}
	\end{equation}
	and the logical operators are given by $\bar{X} = X_3 X_5 X_7$ and $\bar{Z} = Z_1 Z_5 Z_9$.
	
	\begin{figure}
		\centering
		\includegraphics[width=0.99\columnwidth]{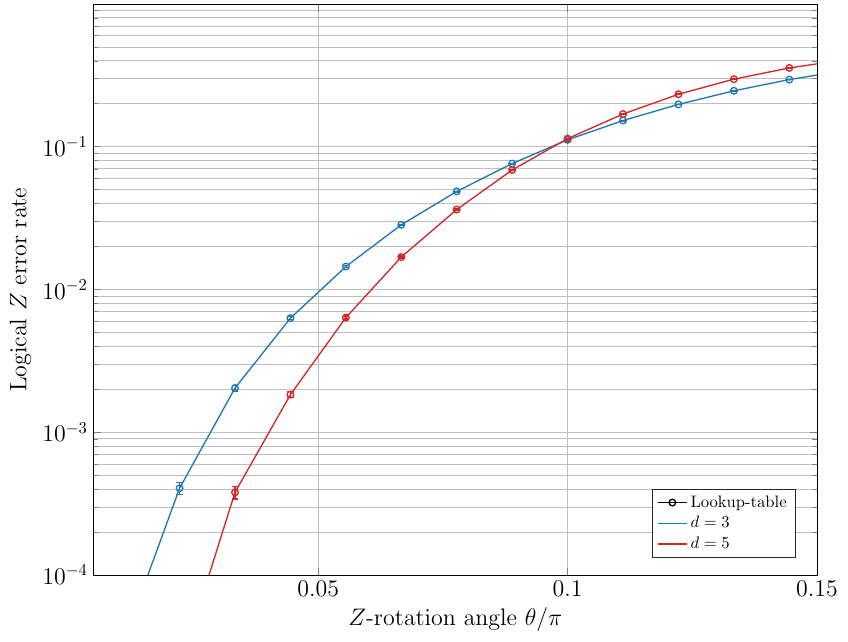}
		{
			\caption{Logical $Z$ error rates of $[[d^2,1,d]]$ rotated surface codes with $d=3$ and $d=5$ under a layer of Pauli-twirled coherent rotation errors. The Pauli-twirled channel is equivalent to independent $Z$ errors with probability $\sin^2\theta$. The horizontal axis denotes the coherent rotation angle $\theta/\pi$.}
			\label{fig:surface_twirled}
		}
	\end{figure}
}

	Figure~\ref{fig:surface_twirled} reproduces this behavior for rotated surface codes with $d=3$ and $d=5$ using a minimum-weight perfect matching decoder~\cite{HG25}.
The crossing point of the two logical error-rate curves indicates a threshold rotation angle of approximately $0.1\pi$, consistent with the result reported in~\cite{BEK+18}.

\begin{figure}[htbp]
	\centering
	\includegraphics[width=1\linewidth]{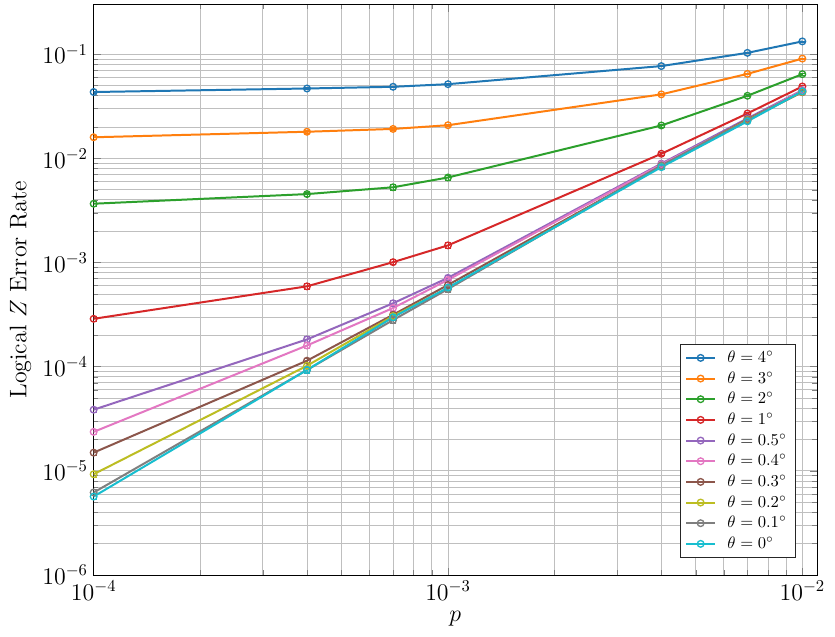}
	\caption{Logical $Z$ error rates of the $[[9,1,3]]$ rotated surface code for different coherent rotation angles $\theta$.}
\label{fig:913_resultp}
\end{figure}

{
However, a fault-tolerant error-correction cycle accumulates many layers of CC errors together with circuit-level noise.
As a result, coherent errors can build up throughout the syndrome extraction process, leading to a substantial degradation in logical error-rate performance even when the rotation angle is relatively small.

Figure~\ref{fig:913_resultp} shows that the logical $Z$ error-rate curves shift upward as the coherent rotation angle $\theta$ increases. For sufficiently large $\theta$, the curves become nearly flat, indicating that coherent errors dominate the decoding performance.

}

\subsubsection{CE CSS codes }

In this section, we develop an efficient simulation procedure for circuit-level noise with interleaved layers of CC errors acting on CE CSS codes.
{
We first introduce a compact representation of the circuit evolution, which allows both Pauli errors and coherent rotations to be tracked efficiently throughout the computation.
This representation forms the basis of Algorithm~\ref{alg:ext_stabilizer}.

The input circuit $\mathcal{C}$ is viewed as a sequence of ideal circuit operations interleaved with fault events and layers of CC errors.
At each time step, the circuit element is either an ideal gate or measurement, or an injected fault event consisting of a Pauli error or a coherent rotation.
The simulation proceeds by maintaining an accumulated Pauli frame together with a set of residual coherent rotations, which are updated according to the propagation rules of the ideal circuit.

	We represent the state in a normal form consisting of a Pauli frame together with a collection $\cE$ of coherent rotations:
	\begin{align}
		\left( \prod_{(\theta, \bfc)\in\cE}e^{i\theta Z^{\bfc}} \right) X^{\bfa} Z^{\bfb} \left( \ket{\phi}  \frac{1}{\sqrt{2}}\left(  \ket{01}^{\otimes w}+ \ket{10}^{\otimes w} \right)\right), \label{eq:representation}
	\end{align}
	where $\ket{\phi}$ is an encoded state. We adopt the convention that all coherent rotations are placed to the left of the Pauli frame.
	The Pauli frame is initialized to $\bfa=0^N$, $\bfb=0^N$, and $\cE=\emptyset$. 
}

Our modified Shor syndrome extraction circuits in Algorithm~\ref{alg:modified_shor}
implement stabilizer measurements while preserving the CE codespace.
Moreover, the codespace and the ancillary qubits are designed to be in the eigenspace of the CC errors. 
The overall effects of the CNOT gates and  CC errors result in  zero error syndromes. 
As a result, we can simulate the evolution of Pauli errors and non-collective coherent rotations through the circuits.

When a new Pauli error is introduced, it is propagated through existing coherent rotations via conjugation rules in Lemma~\ref{lemma:CE_identity}, resulting in deterministic updates of the phase of the coherent rotations.

When a layer of CC errors is introduced, we commute the CC errors through the Pauli frame.
According to Lemma~\ref{lemma:CE_identity}, the resulting effects can be represented as coherent rotations induced by Pauli-$X$ errors.
{
	In particular, when an $X$ error propagates through a layer of CC errors with phase $\theta$, it induces a single-qubit coherent rotation with phase $-2\theta$ by Lemma~\ref{lemma:CE_identity}. These single-qubit coherent rotations need to be recorded. 
}

Therefore, the nonzero $X$ part in the Pauli frame will induce single-qubit coherent rotations in the circuits
and this is the only source of single-qubit coherent rotations.
Single-qubit coherent rotations may propagate through two-qubit gates into two-qubit coherent rotations and one of the two qubits involved in each two-qubit gate is an ancilla qubit that is immediately measured.
Therefore,  coherent rotations throughout the circuits have support on at most two qubits.

The ancillary qubits are measured in the $X$ basis immediately after the transversal two-qubit gates and the corresponding CC error layer.

{
\textbf{Lemma}  \ref{lemma:twirling_measurement}. (restated)
Let $g$ be the stabilizer measured by the modified Shor syndrome extraction circuit, and suppose the input state is
\[
e^{i\theta Z^{\bfc}}E\ket{\phi},
\]
where $\ket{\phi}$ is a code state, $E=X^{\bfa}Z^{\bfb}$ is the current Pauli frame, and
$Z^{\bfc}$ is a coherent rotation.

If $Z^{\bfc}$ commutes with $g$, then the coherent rotation does not affect the syndrome measurement and remains unchanged after the measurement.

If $Z^{\bfc}$ anticommutes with $g$, then the syndrome measurement projects the state to
\[
\begin{cases}
	E\ket{\phi}, & \text{with probability }\cos^2\theta,\\[1ex]
	Z^{\bfc}E\ket{\phi}, & \text{with probability }\sin^2\theta,
\end{cases}
\]
and the measured syndrome bit is flipped in the second case.
}

{
\begin{proof}
	The modified Shor syndrome extraction measures the projector
	$
	P_\pm(g)=\frac{I\pm(-1)^cg}{2},
	$
	where the sign depends on the current Pauli frame.
	
	If $Z^{\bfc}$ commutes with $g$, then
	\[
	P_\pm(g)e^{i\theta Z^{\bfc}}
	=
	e^{i\theta Z^{\bfc}}P_\pm(g),
	\]
	so the coherent rotation does not alter the syndrome measurement and simply remains after projection.
	
	Now suppose $Z^{\bfc}$ anticommutes with $g$.
	Using the identity $	e^{i\theta Z^{\bfc}}
	=
	\cos\theta\,I
	+i\sin\theta\,Z^{\bfc},
	$ and $P_+(g)Z^{\bfc}=	Z^{\bfc}P_-(g),$
	we obtain
	On the other hand, suppose $Z^bfc$ anticommutes with $g$. 
	For convenience, let $\ket{\textrm{even}}=\sum_{v\in\{0,1\}^{2w}: \atop \textrm{wt}(v) \textrm{ even}}  \ket{v}$,
	and $\ket{\textrm{odd}}= \sum_{v\in\{0,1\}^{2w}: \atop \textrm{wt}(v)=1\mod 2}  \ket{v}$, where $w$ can be inferred from the context and is omitted.	
	Following the deduction in (\ref{eq:modified_shor}), we have
	\begin{align*}
		& P_+(g)\ket{\psi} \ket{\textrm{even}}+  P_-(g)\ket{\psi} \ket{\textrm{odd}}\\
		=&\cos\theta  P_+(g)E \ket{\phi} \ket{\textrm{even}}+  i\sin\theta Z^{\bfc} P_-(g)E \ket{\phi} \ket{\textrm{even}}\\
		+&\cos\theta  P_-(g)E \ket{\phi} \ket{\textrm{odd}}+  i\sin\theta Z^{\bfc} P_+(g)E \ket{\phi} \ket{\textrm{odd}}.
	\end{align*}
	
	Depending on whether the current Pauli frame $E$ commutes or anticommutes with $g$, the above state reduces to one of the following forms:
	\[
	\cos\theta P_+(g)E\ket{\phi}\ket{\textrm{even}}
	+i\sin\theta Z^{\bfc}P_+(g)E\ket{\phi}\ket{\textrm{odd}},
	\]
	or
	\[
	\cos\theta P_-(g)E\ket{\phi}\ket{\textrm{odd}} +i\sin\theta Z^{\bfc}P_-(g)E\ket{\phi}\ket{\textrm{even}}.
	\]
	In both cases, measuring the ancilla in the computational basis yields
	\[
	\begin{cases}
		E\ket{\phi}, & \text{with probability }\cos^2\theta;\\[1ex]
		Z^{\bfc}E\ket{\phi}, & \text{with probability }\sin^2\theta,
	\end{cases}
	\]
	where the latter outcome also flips the syndrome bit.
	
Therefore, after the syndrome measurement, the coherent rotation can be replaced by a stochastic Pauli error $Z^{\bfc}$ occurring with probability $\sin^2\theta$, and the residual coherent rotation is removed.
\end{proof}
}

{
The ideal operations considered in Algorithm~\ref{alg:ext_stabilizer} consist of controlled-$X/Z$ gates, zero-controlled-$X/Z$ gates, single-qubit Pauli operations, and syndrome measurements.
For the distance-three CE CSS codes studied in this work, the algorithm exactly captures the dominant single- and double-fault coherent error events, while higher-order fault events involving three or more coherent interactions are approximated by the Pauli-twirling procedure described above.

By Lemma~\ref{lemma:twirling_measurement}, every coherent rotation whose support overlaps the measured ancilla or data qubit is collapsed into an equivalent stochastic Pauli error with probability $\sin^2\theta$.
Therefore, after each ancilla measurement, all coherent rotations involving the measured ancilla are removed from the coherent error set $\mathcal{E}$.
}

During each ancilla measurement, every coherent rotation involving the measured ancilla is projected according to Lemma~\ref{lemma:twirling_measurement}. The resulting Pauli error is absorbed into the Pauli frame, and the corresponding coherent rotation is removed from $\mathcal{E}$.

{
For the distance-three CE CSS codes considered in this work, the modified Shor syndrome extraction circuits use only transversal two-qubit gates. Therefore, single- and double-fault coherent error events generate coherent rotations with support on at most two data qubits before the subsequent syndrome extraction round.
Consequently, the resulting coherent error on the data qubits can be expressed as
\[
\alpha_0 I+\alpha_1 Z_i+\alpha_2 Z_j+\alpha_3 Z_iZ_j .
\]
Since the code distance is three, none of these weight-one or weight-two $Z$-type operators can be a logical operator. The subsequent complete syndrome extraction round projects different Pauli components into distinct syndrome subspaces, thereby eliminating the interference between them. 

Consequently, the effect of these coherent rotations is equivalent to independently applying the Pauli-twirled approximation in Lemma~\ref{lemma:twirling_measurement} to each coherent rotation and combining the resulting Pauli errors. Therefore, Algorithm~\ref{alg:ext_stabilizer} exactly captures all dominant single- and double-fault coherent error events.

The only approximation in Algorithm~\ref{alg:ext_stabilizer} arises from higher-order fault events involving three or more coherent error events, which  may generate coherent superpositions involving multiple Pauli components with larger support. 
Tracking such coherent superpositions exactly would require maintaining the amplitudes of exponentially many Pauli components, which is beyond the stabilizer formalism.

Therefore, during syndrome measurements, we apply Lemma~\ref{lemma:twirling_measurement}, while neglecting interference among different Pauli components. 

For the final perfect syndrome extraction round used only for determining the logical error rate, we assume that Pauli twirling is applied independently to each qubit before the syndrome measurement.

Consequently, Algorithm~\ref{alg:ext_stabilizer} exactly captures the dominant fault events while providing an efficient approximation for higher-order coherent interactions.
}

{
\subsection{Proof of Weak FTEC in Theorem~\ref{thm:weak_FTEC}}

	We verify the two conditions of weak FTEC in Def.~\ref{def:weak_FTEC}.
	
	First, we consider the correction condition. 
	For the CE CSS codes considered in this work, the modified Shor syndrome extraction circuits consist only of transversal two-qubit gates between data and ancilla qubits, followed immediately by $X$-basis measurements of the ancilla qubits.
	Hence, coherent rotations generated during syndrome extraction cannot propagate beyond a single data--ancilla pair before the corresponding ancilla measurement.
	Therefore, it suffices to consider coherent rotations of the form
	\[
	e^{i\theta Z_{(d)}Z_{(a)}} .
	\]
	
	A Pauli $X$ component in the Pauli frame may interact with a subsequent CC error layer and induce a single-qubit coherent rotation on the same qubit according to Lemma~\ref{lemma:CE_identity}.
	This is the only source of single-qubit coherent rotations in the circuit evolution.
	Since such coherent rotations are in the $Z$ direction, they commute with $Z$-type stabilizer measurements and do not affect the corresponding syndrome outcomes.
	Therefore, any residual $X$ error component must be removed before performing the final $Z$-type correction.
	
	We now analyze the control flow in Algorithm~\ref{alg:control_flow}.
	If the first round of $Z$-type stabilizer measurements gives a nonzero outcome, the outcome may originate from either a data error or a fault occurring during syndrome extraction.
	Therefore, a second round of $Z$-type stabilizer measurements is performed, and the resulting syndrome is used to determine the $X$ correction.
	After removing the $X$ component, the subsequent $X$-type stabilizer measurements determine the remaining $Z$ correction.
	During the $X$-type stabilizer measurements, coherent rotations may be projected into stochastic $Z$ errors according to Lemma~\ref{lemma:twirling_measurement}.
	In particular, a rotation $e^{i\theta Z_{(d)}Z_{(a)}}$ may induce a $Z$ error on the data qubit and flip the measured syndrome bit.
	Hence, a second round of $X$-type stabilizer measurements is required before applying the final $Z$ correction.
	
	If the first round of $Z$-type stabilizer measurements gives a zero outcome, Algorithm~\ref{alg:control_flow} proceeds to measure the $X$-type stabilizers.
	A nonzero outcome of the first $X$-type syndrome measurement may correspond to an existing $Z$ error or the $Z$ component of an  $Y$ error.
	In the latter case, the residual $X$ component may generate coherent rotations under subsequent CC error layers.
	Therefore, an additional $Z$-type syndrome measurement is performed to remove the $X$ component before the final $X$-type syndrome measurement and $Z$ correction.
	
	For distance-three CE CSS codes, the above procedure corrects every single-fault event.
	Thus the error correction condition of Def~\ref{def:weak_FTEC} is satisfied.
	
	It remains to verify the containment condition.
	The modified Shor syndrome extraction circuits contain only transversal two-qubit gates, and thus a single fault cannot propagate to multiple data qubits.
	Moreover, a single fault involving a CC error layer can only generate a coherent rotation supported on the same data qubit where the fault occurs before the corresponding syndrome measurement. Therefore, after the error correction procedure, the resulting data error has support on at most the original faulty data qubit, satisfying the containment condition of Def.~\ref{def:weak_FTEC}.
}

{
\subsection{Syndrome extraction circuits for the $[[14,1,3]]$ CE CSS code}
\label{app:1413}

%


As discussed in Appendix~\ref{app:constructions},  applying Theorem~\ref{thm:construction} to the $[[7,1,3]]$ Steane code~\cite{Ste96} yields a $[[14,1,3]]$ CE CSS code.

The stabilizer generators are scheduled in six measurement depths, as shown below. Stabilizers separated by a horizontal line are measured in different depths, whereas stabilizers within the same block have disjoint supports and can therefore be measured simultaneously. Here,  $g_5'=g_5 g_9 g_{13} $ is chosen so that $g_4$ and $g_5'$ can be measured in the same depth.

{
	\begin{table}[H]
		\centering
		\begin{tabular}{ccccccccccccccc}
			$g_4 =$& $Z$ &$I$ &$I$ &$I$ &$Z$ &$I$ &$I$ &$I$ &$Z$ &$I$ &$I$ &$I$ &$Z$ &$I$ \\
			$g_5'=$ &$I$&$I$&$Z$&$I$&$I$&$Z$&$I$&$I$&$I$&$I$&$Z$&$I$&$I$&$Z$ \\
			\hline
			$g_6=$ &$I$&$I$&$I$&$I$&$I$&$I$&$Z$&$I$&$Z$&$I$&$Z$&$I$&$Z$&$I$ \\
						$g_7 =$&$-Z$ &$Z$ &$I$ &$I$ &$I$ &$I$ &$I$ &$I$ &$I$ &$I$ &$I$ &$I$ &$I$ &$I$ \\
			$g_8 =$& $-I$ &$I$ &$Z$ &$Z$ &$I$ &$I$ &$I$ &$I$ &$I$ &$I$ &$I$ &$I$ &$I$ &$I$ \\
			$g_9 =$& $-I$ &$I$ &$I$ &$I$ &$Z$ &$Z$ &$I$ &$I$ &$I$ &$I$ &$I$ &$I$ &$I$ &$I$ \\
			\hline
			$g_{10} =$& $-I$ &$I$ &$I$ &$I$ &$I$ &$I$ &$Z$ &$Z$ &$I$ &$I$ &$I$ &$I$ &$I$ &$I$ \\
			$g_{11} =$& $-I$ &$I$ &$I$ &$I$ &$I$ &$I$ &$I$ &$I$ &$Z$ &$Z$ &$I$ &$I$ &$I$ &$I$ \\
			$g_{12}=$& $-I$ &$I$ &$I$ &$I$ &$I$ &$I$ &$I$ &$I$ &$I$ &$I$ &$Z$ &$Z$ &$I$ &$I$ \\
			$g_{13}=$& $-I$ &$I$ &$I$ &$I$ &$I$ &$I$ &$I$ &$I$ &$I$ &$I$ &$I$ &$I$ &$Z$ &$Z$\\
			\hline
			$g_1 =$& $X$ & $X$ &$I$ &$I$ & $X$ & $X$ & $I$ & $I$ &$X$ &$X$ & $I$ & $I$ &$X$ &$X$ \\
			\hline
			$g_2 =$& $I$ &$I$ & $X$ & $X$ & $X$ & $X$& $I$ & $I$ & $I$ & $I$ &$X$ &$X$ &$X$ &$X$  \\
			\hline
			$g_3 =$& $I$ &$I$ &$I$ &$I$ &$I$ &$I$ & $X$ & $X$ & $X$ & $X$ & $X$ & $X$ & $X$ & $X$ \\
		\end{tabular}
	\end{table}
}

Using the simulation procedure in Algorithm~\ref{alg:ext_stabilizer}, we demonstrate that the $[[14,1,3]]$ CE Steane code maintains a stable logical error rate under CC errors, as shown in Fig.~\ref{fig:1413_resultp}.

   \begin{figure}[htbp]
	\centering
	\includegraphics[width=1\linewidth]{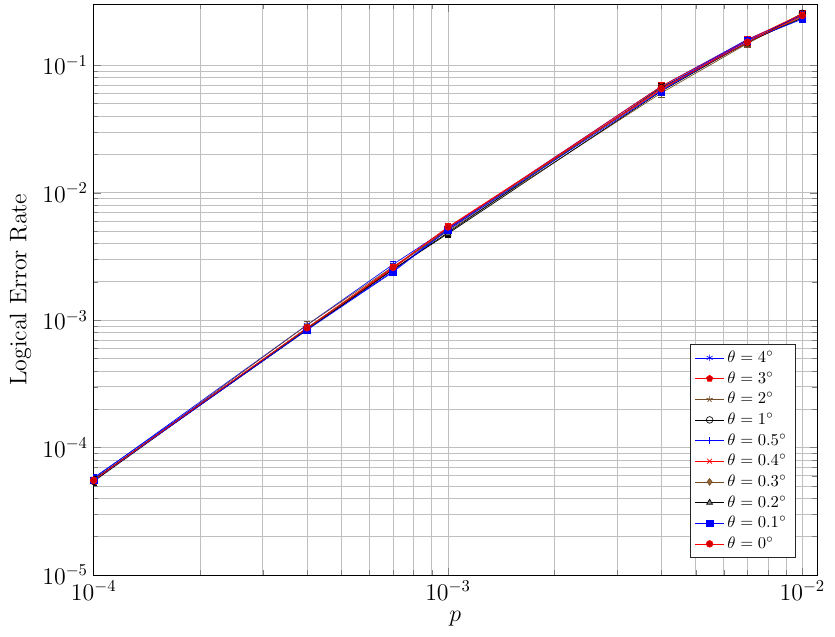}
	\caption{Logical error rates of the $[[14,1,3]]$ CE Steane code for different coherent rotation angles $\theta$.} \label{fig:1413_resultp}
\end{figure}
}     
     
\end{document}